\documentclass[aps,twocolumn,prb,superscriptaddress]{revtex4-2}

\usepackage{graphicx} 
\usepackage{amsmath}
\usepackage{amssymb}
\usepackage[colorlinks=true]{hyperref}
\hypersetup{allcolors=blue}
\usepackage{array}
\usepackage{multirow}
\usepackage{xcolor}
\usepackage{lipsum}
\usepackage{appendix}
\usepackage{subcaption}
\usepackage{mhchem}
\usepackage{orcidlink}
\usepackage{placeins}
\usepackage{capt-of}
\usepackage{natbib}
\newcommand\Tstrut{\rule{0pt}{2.6ex}}         % = `top' strut
\begin{document}

%\title{Non-universality in the physics of higher-order Van Hove singularities}

%\title{Breakdown of the hot-spot approximation at higher-order Van Hove singularities}

%\title{How universal is higher-order-Van-Hove-singularity physics in reality?}

%\title{Manipulating correlated ground states using higher-order van-hove singularities}

\title{Engineering correlated phases through manipulation of Van Hove singularities}

%\title{Competing Orders from Higher-Order Van Hove Singularities}

%\title{Influence of higher-order Van Hove singularities on the correlated ground state of the square lattice Hubbard model}
%Just a dummy title - please edit!

\author{Thomas P. Sheerin\,\orcidlink{0000-0001-5015-0769}}
\affiliation{SUPA, School of Physics and Astronomy, University of St Andrews,
North Haugh, St Andrews, Fife KY16 9SS, United Kingdom}

\author{Maria Ramirez\,\orcidlink{0009-0000-5019-4008}}
\affiliation{SUPA, School of Physics and Astronomy, University of St Andrews,
North Haugh, St Andrews, Fife KY16 9SS, United Kingdom}

\author{Chris A. Hooley\,\orcidlink{0000-0002-9976-2405}}
\affiliation{Centre for Fluid and Complex Systems, Coventry University, Coventry CV1 2TT, United Kingdom}

\author{Luke C. Rhodes\,\orcidlink{0000-0003-2468-4059}}
\email[]{lcr23@st-andrews.ac.uk}
\affiliation{SUPA, School of Physics and Astronomy, University of St Andrews,
North Haugh, St Andrews, Fife KY16 9SS, United Kingdom}

\begin{abstract}
Controlling the ordered phases of correlated electron systems remains a central challenge in quantum materials design. Divergences in the electronic density of states, known as Van Hove singularities (VHSs), are one obvious route to such control.  It is clear from recent work that the exact functional form of these divergences can profoundly affect which phases are realized; a full picture, however, remains elusive. In this work, we use both the hot-spot parquet renormalization group and the truncated-unity functional renormalization group to theoretically study the emergent correlated states of a two-dimensional square-lattice Hubbard model with VHSs at or near the Fermi level. By varying a single hopping parameter, $t_3$, we are able to change  the strength of the VHS divergence in the density of states from logarithmic (for $t_3 < t_{3c}$) to power-law (for $t_3 = t_{3c}$).  Further increase of $t_3$ ($t_3 > t_{3c}$) causes each original Van Hove point to split into two, both of the conventional logarithmic type.  We show that which of these regimes we are in strongly influences the predicted ordered states.  We also study the dependence on doping, and find that the ferromagnetic state that occurs at Van Hove filling in these models is unstable to very small shifts in the Fermi level, often giving way to distinct ordered states depending on whether the model is electron- or hole-doped. These results highlight the importance of tuning VHS properties to control ordered states in correlated materials, and offer design rules to engineer these phases in novel systems. 
\end{abstract}

\maketitle

%==============================
%==============================
%Paper proper:
%==============================
%==============================
\allowdisplaybreaks
\section{Introduction}
\label{sec:intro}

Saddle-point Van Hove singularities (VHSs) in two-dimensional materials have long been recognized as playing an integral role in determining the systems' ground-state properties. The associated logarithmic divergence of the density of states (DOS) at the energy of the singularity can, if tuned to the Fermi level, modify transition temperatures~\cite{Steppke_Strong_2017} and 
stabilize exotic states~\cite{JSJTGW2007,GLALJL2010,Benhabib_Collapse_2015,XWTSTM2021,CMPMRF2024}.
In recent years, much attention has been devoted to higher-order Van Hove singularities (HOVHSs) \cite{Classen_HighOrder_2025}, which correspond to points in the Brillouin zone at which both a band's gradient and the determinant of its Hessian matrix vanish.  This increased flatness of the dispersion creates stronger, power-law singularities in the DOS, which have been suggested to further dramatically alter the system's low-energy properties~\cite{JMABTO2010,Efremov_Multicritical_2019,LCACCH2020,Classen_HighOrder_2025}. Understanding the influence of ordinary and higher-order VHSs on materials' ground-state phase diagrams is not only important from the viewpoint of fundamental physics; it also continues to play an ever more important role in the design of correlated materials~\cite{Chandrasekaran_Engineering_2024}. 

The renormalization group (RG) is a powerful approach to exploring how such changes in fermiology influence the material's low-temperature phase behavior. The hot-spot parquet renormalization group (pRG) is one RG method that has in the past been extensively used to study finite-density fermionic systems at weak coupling~\cite{HS1987,PLGMDP1987,NFTR1998,NFTRMS1998,Irkhin_2001,KLTR2009,RNLLAC2012,SMAC2013,SWSS2014,XCYYHY2015,HYFY2015,JHCHHL2016,ASGGCC2017,YSJB2018,WQLLZZ2019,MTCH2020,LCACCH2020,MTCH2021,ZWYWFW2023,XHASXW2023,AZDEJB2023,Lee_Unified_2025}. This approach proceeds on the assumption that only a few points on the Fermi surface (such as Van Hove points or nested sections) drive the low-energy physics; it works by taking account of only small ``patches'' of momentum space around these points. The possibility of error this introduces is often regarded as compensated for by the resultant tractability of the calculation, and its potential for analytic insight.

By contrast, starting from the Wetterich equation of the functional renormalization group (FRG) and implementing various approximations (such as truncating the effective action at four-fermion terms and assuming static vertices), one can derive weak-coupling RG equations that track the flow of the fully momentum-resolved four-point function, $\Gamma^\Lambda(\mathbf{k}_1,\mathbf{k}_2,\mathbf{k}_3)$, as the RG scale $\Lambda$ is lowered~\cite{MSCH2001,WMMSCH2012}. These equations can in principle be integrated --- however, it is only in recent years that this has become computationally viable. In particular, the development of the truncated-unity functional renormalization group (TUFRG) has significantly reduced the computational cost of this form of FRG, enabling a more thorough exploration of correlated phase spaces \cite{Lichtenstien_TUFRG_2017,JL2018,Profe_TUFRG_2022,Beyer_Reference_2022,Profe_diverge_2024}.

These techniques, pRG and TUFRG, are similar in various ways. In particular, both take account of all one-loop corrections to the four-point function, thereby ensuring that competition between particle-particle and particle-hole channels is correctly treated at weak coupling. In fact, the only respect in which they significantly differ is that TUFRG accounts for much more of the material's band structure than pRG. This fact should presumably make TUFRG the more accurate method. It therefore seems timely to consider how the predictions of the two methods compare.

\begin{figure*}
\begin{center}
\includegraphics[width=0.8\linewidth]{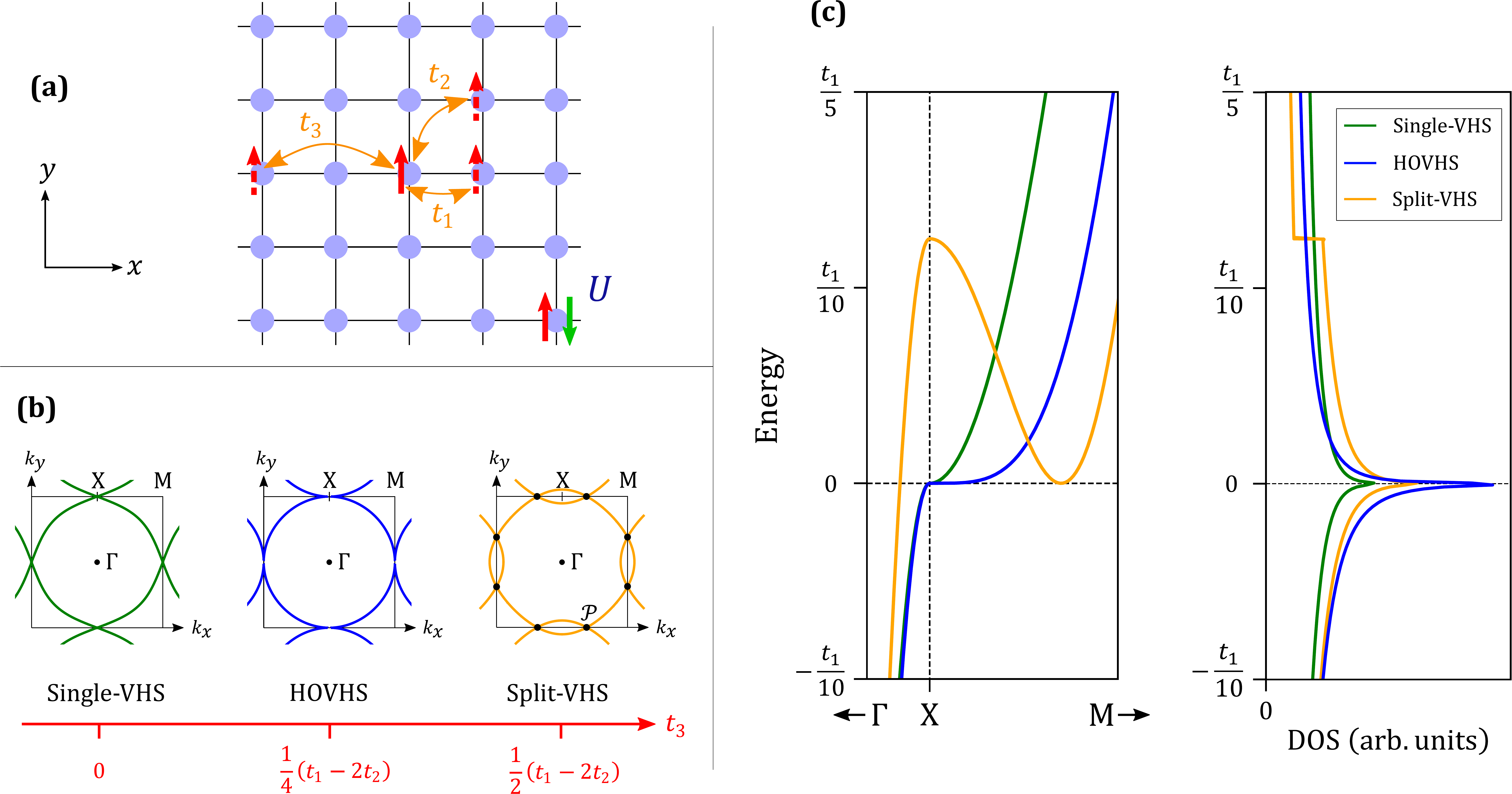}
\end{center}
\caption{(a) A real-space sketch of the square-lattice Hubbard model that we study. The model has first-, second-, and third-neighbor hopping terms with amplitudes $t_1$, $t_2$, and $t_3$ respectively, as well as an on-site Coulomb repulsion term with strength $U$.  The Hamiltonian of the model is given in (\ref{eqn:Hubbard_realspace}). (b) Non-interacting Fermi surfaces of the model, with the Fermi level set to the energy of the VHSs, in three regimes:\ single-VHS ($t_3 < t_{3c}$), higher-order VHS ($t_3 = t_{3c}$), and split-VHS ($t_3 > t_{3c}$).  Here $t_{3c} = (t_1-2t_2)/4$. (c) The band structure near the X point, and the density of states near the Fermi level, in the three regimes. Note the enhanced divergence in the density of states at the Fermi level in the HOVHS case. In (b) and (c) we have set $t_2/t_1=0.375$ (other choices give similar plots).}
\label{fig:fig1}
\end{figure*}

In this paper we compare the predictions of the pRG and TUFRG methods for the square-lattice Hubbard model.  We include first-, second-, and third-neighbor hopping processes; tuning the amplitude of the third-neighbor hopping ($t_3$) allows us to vary the type and number of Van Hove points in the non-interacting band structure.  When we fix the filling so that the VHSs occur at the Fermi level, we find that the weak-coupling phase diagram depends smoothly but highly non-trivially on $t_3$, i.e.\ the type and number of the VHSs strongly influences the predicted correlated ground state.

We then consider the effect of doping the system away from Van Hove filling.  We find significantly more sensitivity to this perturbation, with ferromagnetic phases disappearing when the Fermi level is moved away from the VHSs by very small fractions of the bandwidth. The phases obtained slightly away from Van Hove filling (which include triplet superconductivity and an incommensurate spin-density wave) are more robust to changes in the Fermi level, but are nevertheless affected by the proximity of the VHS. In particular, in the case of a HOVHS the asymmetry of the DOS singularity means that different correlated ground states are stabilized depending on whether the VHS is occupied or unoccupied, i.e.\ on whether we dope with electrons or holes. 

The paper is organized as follows. In section~\ref{sec:model} we briefly outline the third-nearest-neighbor Hubbard model and the VHSs of its non-interacting band structure. In section~\ref{sec:setup} we detail the renormalization-group procedures we use to study the model. We present our results in section~\ref{sec:Results}, beginning with the phase diagrams of the system when it is at Van Hove filling (section~\ref{subsec:results_VHfilling}) and then exploring how the ordered state varies with filling (section~\ref{subsec:doping}). We conclude in section~\ref{sec:discussion} with some comment on the shortcomings of hot-spot renormalization-group schemes, as well as discussion of the implications of our results for the project of using VHSs for quantum materials design.

\section{Model}
\label{sec:model}

The Hamiltonian we consider is
\begin{align}
\label{eqn:Hubbard_realspace}
&\hat{H}=-\sum_\sigma\sum_{i,j}t_{ij}\,\hat{c}^\dag_{i,\sigma}\hat{c}_{j,\sigma}+U\sum_{i}\hat{n}_{i,\uparrow}\hat{n}_{i,\downarrow}~,
\end{align}
%Here, $\hat{H}$ is the Hamiltonian, $\hat{N}$ is the number operator, 
where the hopping parameters $t_{ij}$ are real and symmetric under $i\leftrightarrow j$, and $U$ is the strength of the on-site Coulomb repulsion.  The operator $\hat{c}^\dag_{i,\sigma}$ creates an electron of spin projection $\sigma \in \{\uparrow,\downarrow\}$ on site $i$, and $\hat{n}_{i,\sigma}=\hat{c}^\dag_{i,\sigma}\hat{c}_{i,\sigma}$. %We now Fourier-transform and take the thermodynamic limit, getting
%\begin{align}
%\label{eqn:Hubbard_kspace}
%&\hat{H}-\mu \hat{N}=\sum_{\sigma}\int_{\mathbf{k}}\xi(\mathbf{k})\hat{\psi}^\dag_\sigma(\mathbf{k})\hat{\psi}_\sigma(\mathbf{k})\nonumber\\&+\frac{U}{2}\sum_{\sigma,\sigma'}\int_{\mathbf{k}_1,\mathbf{k}_2,\mathbf{k}_3}\hat{\psi}^\dag_\sigma(\mathbf{k}_1)\hat{\psi}^\dag_{\sigma'}(\mathbf{k}_2)\hat{\psi}_{\sigma'}(\mathbf{k}_3)\hat{\psi}_{\sigma}(\mathbf{k}_4)~,
%\end{align}
%where we have set the lattice spacing to unity and $\hat{\psi}^\dag_{\sigma}(\mathbf{k})$ is an appropriate rescaling of the Fourier transform of $c^\dag_{i,\sigma}$. The momentum $\mathbf{k}_4$ is set by momentum conservation modulo umklapp scattering. 

We take the hopping parameters to include terms up to and including third-neighbor hopping; the resulting non-interacting dispersion relation, $\xi(\mathbf{k})$, is
\begin{align}
\label{eqn:disp}
\xi(\mathbf{k})=&-2t_1\left[\cos(k_x)+\cos(k_y)\right]\nonumber\\ &-4t_2\cos(k_x)\cos(k_y)\nonumber\\ &+2t_3\left[\cos(2k_x)+\cos(2k_y)\right] -\mu~,
\end{align}
in units in which the lattice spacing is unity. The energy offset $\mu$ is chosen so that the Van Hove singularities occur at energy $\xi = 0$ (in section~\ref{subsec:doping} we add a further energy shift $\Delta E_\text{VHS}$ in order to move the Van Hove singularities away from the Fermi level). Throughout this paper we assume that $t_1>2t_2>0$. We have adopted the convention that $t_3$ appears in (\ref{eqn:disp}) with a `+' sign before it, and so the regime of interest (i.e.\ the one in which a HOVHS can be achieved) is $t_3\geqslant 0$.
%In order to reach the HOVHS electronic structure, $t_3$ must enter with opposite sign to $t_1$ and $t_2$, we thus restrict ourselves to and $t_3\geqslant0$ and include the opposite sign in Eq. \eqref{eqn:disp}. 
We present all results in terms of the dimensionless ratios $t_2/t_1$, $t_3/t_1$ and $U/t_1$. 

A sketch of the real-space system is presented in Fig.~\ref{fig:fig1}(a).  The square lattice results in a square Brillouin zone with high-symmetry points $\Gamma=(0,0)$, $\text{X}=(\pi,0),(0,\pi)$, and $\text{M}=(\pi,\pi)$, shown in Fig.~\ref{fig:fig1}(b).  Following usual convention, we have specified here only those high-symmetry points that are not related to each other by reciprocal lattice vectors.

Tuning $t_3$ allows us to vary the functional form of the divergence in the density of states of the VHS, as well as the positions of the Van Hove points on the Brillouin-zone boundary, as shown in Figs.~\ref{fig:fig1}(b) and \ref{fig:fig1}(c). At $t_3=0$ there are quadratic saddle points located at X, giving a logarithmic divergence in the DOS at the Fermi level.  Increasing $t_3$ causes the band dispersion along the X-M high-symmetry directions to become increasingly flat, until at $t_3 = t_{3c} \equiv (t_1-2t_2)/4$ the saddle points become of higher order and a HOVHS develops (of the cusp ($A_3$) type~\cite{ACASJB2020}, with DOS $\sim |\xi|^{-1/4}$). For $t_3>t_{3c}$ the saddle point at each X point splits into two quadratic saddle points located on the Brillouin-zone boundary, leaving behind a band maximum at X. These new saddle points are positioned at $\mathcal{P}=(\pi,\pm k_\mathcal{P})$ and $(\pm k_\mathcal{P},\pi)$, with \begin{align}
k_\mathcal{P} = \arccos\left(\frac{t_1-2t_2}{4t_3}\right)~,
\end{align}
and give rise to a logarithmic divergence in the DOS at $\xi = 0$. There is further a step-like discontinuity at a positive value of $\xi$ arising from the band maxima at the X points.

There are thus three qualitatively distinct regimes of the model:\ the single-VHS case, $t_3 < t_{3c}$; the HOVHS case, $t_3 = t_{3c}$; and the split-VHS case $t_3 > t_{3c}$.  As we shall see, which of these regimes we are in has a profound effect on the types of correlated low-temperature state that occur when the effects of electron-electron interactions are taken into account.

\section{Renormalization-group frameworks}
\label{sec:setup}

\subsection{Hot-spot parquet renormalization group (pRG)}
\label{subsec:prg}

When there are VHSs at the Fermi level, it is quite common to assume that the low-energy physics is dictated entirely by the details of the dispersion in the immediate vicinity of the associated Van Hove points in the Brillouin zone. This is the hot-spot approximation --- it enables a pRG analysis that is mostly analytically tractable, and so has in the past given rise to valuable insight into many materials~\cite{HS1987,PLGMDP1987,NFTR1998,NFTRMS1998,Irkhin_2001,KLTR2009,RNLLAC2012,SMAC2013,SWSS2014,XCYYHY2015,HYFY2015,JHCHHL2016,ASGGCC2017,YSJB2018,WQLLZZ2019,MTCH2020,LCACCH2020,MTCH2021,XHASXW2023,ZWYWFW2023,AZDEJB2023,Lee_Unified_2025}.

\begin{figure}
\centering
\includegraphics[width=0.9\linewidth]{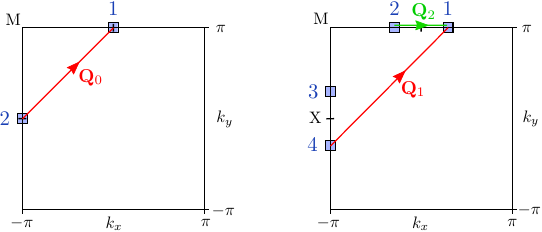}
\caption{The two-patch scheme used in the pRG analysis of the single-VHS and HOVHS cases (left), and the four-patch scheme used in the split-VHS case (right). Also shown are the symmetry-inequivalent inter-patch wavevectors.}
\label{fig:pRG_patching}
\end{figure}

In the pRG analysis presented here, we retain the dispersion only in small patches around the Van Hove points:\ these patches have side length $2k_\text{cut}$, where $k_\text{cut}$ is an ultraviolet momentum cutoff. The patch schemes used are illustrated in Fig.~\ref{fig:pRG_patching}. In the single-VHS and HOVHS cases, the singularities are at $(0,\pi)$ and $(-\pi,0)$. The lattice and point-group  symmetries imply that the only non-zero inter-patch wavevector one has to consider is (say) $\mathbf{Q}_0=(\pi,\pi)$; the susceptibilities at all other inter-patch wavevectors are the same. In the split-VHS case there are four inequivalent patches, and there is a minimal set of two non-zero inter-patch wavevectors, as also shown in Fig.~\ref{fig:pRG_patching}. Note that the wavevector connecting patches 3 and 2, $\mathbf{K}_2-\mathbf{K}_3$, is equal to $-\mathbf{Q}_1$ up to a reciprocal lattice vector, and $\mathbf{K}_1-\mathbf{K}_3$ is similarly equivalent to a rotation of $\mathbf{Q}_1$.

Setting
\begin{align}
m_+^\text{X}&=\frac{1}{2(t_1-2t_2-4t_3)}~,\nonumber\\
m_-^\text{X}&=\frac{1}{2(t_1+2t_2+4t_3)}~,
\end{align}
the effective masses at the X point at $(0,\pi)$ are $m_x=m_+^\text{X}$ and $m_y=m_-^\text{X}$, and so the dispersion is locally given by
\begin{align}
\xi\approx\frac{k_x^2}{2m_+^\text{X}}-\frac{(k_y-\pi)^2}{2m_-^\text{X}}~.
\end{align}
The effective masses at $(-\pi,0)$ are, by symmetry, $m_x=m_-^\text{X}$ and $m_y=m_+^\text{X}$.
%The effective masses of the saddle point at $\text{X}$ are
%\begin{align}
%m_x^\text{X}&=\frac{1}{2(t_1-2t_2-4t_3)}~,\nonumber\\
%m_y^\text{X}&=\frac{1}{2(t_1+2t_2+4t_3)}~,
%\end{align}
%and so the dispersion is locally given by
%\begin{align}
%\xi\approx\frac{k_x^2}{2m_x^\text{X}}-\frac{(k_y-\pi)^2}{2m_y^\text{X}}~.
%\end{align}

In the split-VHS case, we define
\begin{align}
m_+^\mathcal{P}&=-\frac{2t_3}{(t_1-2t_2-4t_3)(t_1-2t_2+4t_3)}~,\nonumber\\
m_-^\mathcal{P}&=\frac{t_3}{(2t_3+t_2)(t_1-2t_2+4t_3)}~,
\end{align}
and the dispersion about, e.g., $(k_\mathcal{P},\pi)$ is 
\begin{align}
\xi\approx \frac{(k_x-k_\mathcal{P})^2}{2m_+^\mathcal{P}}-\frac{(k_y-\pi)^2}{2m_-^\mathcal{P}}~.
\end{align}
%The effective masses at the points $\mathcal{P}$ in the split-VHS case are
%\begin{align}
%m_x^\mathcal{P}&=-\frac{2t_3}{(t_1-2t_2-4t_3)(t_1-2t_2+4t_3)}~,\nonumber\\
%m_y^\mathcal{P}&=\frac{t_3}{(2t_3+t_2)(t_1-2t_2+4t_3)}~.
%\end{align}
%and the dispersion about, e.g.,\ $(k_\mathcal{P},\pi)$ is 
%\begin{align}
%\xi\approx \frac{(k_x-k_\mathcal{P})^2}{2m_x^\mathcal{P}}-\frac{(k_y-\pi)^2}{2m_y^\mathcal{P}}~.
%\end{align}
The dispersions about the other $\mathcal{P}$ points ($(-k_\mathcal{P},\pi)$ and $(-\pi,\pm k_\mathcal{P})$) are obtained by symmetry. Both $m_+^\text{X}$ and $m_+^\mathcal{P}$ diverge at $t_3=t_{3c}$, corresponding to the development of HOVHSs at the X points. At this value of $t_3$ the dispersion about (for example) $\text{X}=(0,\pi)$ is
\begin{align}
\xi\approx t_3k_x^4+t_2k_x^2(k_y-\pi)^2-\frac{(k_y-\pi)^2}{2m_-^\text{X}}~.
\end{align}

The particle-hole and particle-particle susceptibilities are respectively defined at zero temperature as
\begin{align}
\label{eqn:bubbles}
\Pi_\text{ph}^{\mathbf{q}}(\Omega)&=-\int_{-\infty}^\infty\frac{d\omega}{2\pi}\int_{\mathbf{k}}G(\omega,\mathbf{k})\,G(\omega+\Omega,\mathbf{k}+\mathbf{q})~,\nonumber\\\Pi_\text{pp}^{\mathbf{q}}(\Omega)&=\int_{-\infty}^\infty\frac{d\omega}{2\pi}\int_{\mathbf{k}}G(\omega,\mathbf{k})\,G(\Omega-\omega,\mathbf{q}-\mathbf{k})~.
\end{align}
Here $\Omega$ and $\mathbf{q}$ are a frequency and wavevector, respectively, and $G(\omega,\mathbf{k})=1/(i\omega-\xi(\mathbf{k}))$ is the zero-temperature, non-interacting Matsubara Green's function of the electrons. In a patch scheme, we compute the above only when $\mathbf{q}$ equals one of the inter-patch wavevectors. In addition, $\int_\mathbf{k}=\int d^2\mathbf{k}/(2\pi)^2$, the integration range being restricted to a single patch. In the following, we provide some details of the setup and pRG flow equations, for the single-VHS, HOVHS and split-VHS cases in turn.

\subsubsection{Single-VHS case}

%The single-VHS case has been analysed using pRG methods in the past~\cite{NFTRMS1998} --- however, certain aspects of our treatment are nuanced, and furthermore it mirrors our framework for the HOVHS and split-VHS cases. In all cases,

In the single-VHS case there are four non-interacting susceptibilities to be calculated, namely $\Pi_\text{pp}^{\mathbf{0}}(\Omega)$, $\Pi_\text{ph}^{\mathbf{0}}(\Omega)$, $\Pi_\text{pp}^{\mathbf{Q}_0}(\Omega)$ and $\Pi_\text{ph}^{\mathbf{Q}_0}(\Omega)$. We have~\cite{HYFY2015}
\begin{align}
\label{eqn:prg_log_susc}
\Pi_\text{pp}^{\mathbf{0}}(\Omega)&\sim ~\nu_0\ln^2\frac{W}{|\Omega|}~,&~\Pi_\text{ph}^{\mathbf{0}}(\Omega)&\sim2\nu_0\ln\frac{W}{|\Omega|}~,\nonumber\\\Pi_\text{pp}^{\mathbf{Q}_0}(\Omega)&\sim2\nu_0\gamma_1\ln\frac{W}{|\Omega|}~,&~\Pi_\text{ph}^{\mathbf{Q}_0}(\Omega)&\sim2\nu_0\gamma_2\ln\frac{W}{|\Omega|}~,
\end{align}
where $W$ is a high-energy cutoff of the order of the half-bandwidth of each patch (see appendix~\ref{append:prg_correct_suscs} for discussion), and we have only kept the leading singularities at small $|\Omega|/W$. Throughout this paper, we take $W\approx 0.08\,t_1$ --- this is of the order of the energetic separation between the Fermi level and the maxima at the X points in the split-VHS case (see Fig.~\ref{fig:fig1}(c)), and for simplicity we use the same value in the single-VHS and HOVHS cases. This is quite a simplistic choice, but it at most causes error in the sizes of the Fermi-liquid regions we predict, which transpires not to affect our conclusions. Above, we have furthermore set $\nu_0=\sqrt{m_+^\text{X}m_-^\text{X}}/(4\pi^2)$, and
\begin{align}
\label{eqn:gammas}
\gamma_1&=\frac{2\sqrt{\kappa}}{\kappa-1}\arctan\left(\frac{\kappa-1}{2\sqrt{\kappa}}\right)~,\nonumber\\
\gamma_2&=\frac{2\sqrt{\kappa}}{\kappa+1}\ln\left(\left|\frac{\sqrt{\kappa}+1}{\sqrt{\kappa}-1}\right|\right)~,
\end{align}
with $\kappa=m_+^\text{X}/m_-^\text{X}$. Note that the expressions we use for $\gamma_1$ and $\gamma_2$ differ from those previously used by a number of authors~\cite{HYFY2015,MTCH2020,Lee_Unified_2025}, though match those in Ref.~\cite{Irkhin_2001} when $t_3=0$ --- in appendix~\ref{append:prg_correct_suscs} we present a detailed derivation of our results. Furthermore, we have neglected the part of $\Pi_\text{pp}^\mathbf{0}(\Omega)$ that is proportional to $\ln(W/|\Omega|)$, even though such terms are retained in the other susceptibilities. This approximation is often used in the literature~\cite{NFTRMS1998,KLTR2009,RNLLAC2012,HYFY2015,XCYYHY2015,JHCHHL2016,MTCH2020,ZWYWFW2023}, and we provide some more justification in appendix~\ref{append:prg_correct_suscs}.

\begin{figure}
\centering
\includegraphics[width=0.6\linewidth]{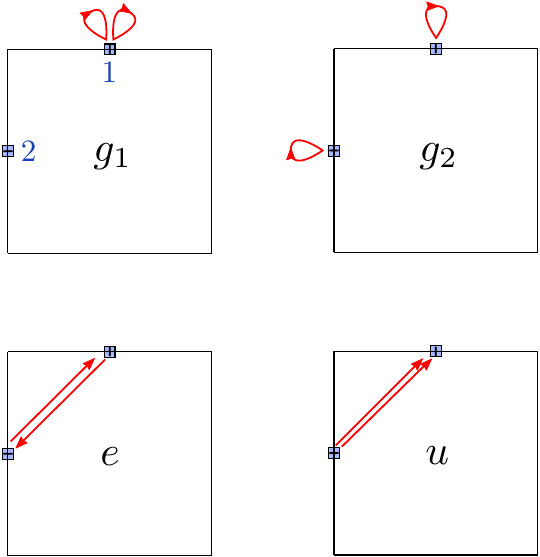}
\caption{The four inequivalent four-point interactions in the two-patch scenarios. The arrows indicate the initial and final momenta of the particles participating in each interaction. $g_1$ and $g_2$ are direct couplings, $e$ is an exchange coupling and $u$ is an umklapp coupling.}
\label{fig:pRG_2patch_scatterings}
\end{figure}

To obtain the renormalization-group flow equations, we first make the following Ansatz for the low-energy Hamiltonian: %Euclidean-time Lagrangian in the continuum:
%\begin{align}
%\label{eqn:Lagrangian}
%\mathcal{L}_{\text{eff}}&=\sum_{a,\sigma}\bar{\psi}_{a\sigma}\bigr(\partial_\tau+\xi^{(a)}(-i\partial_x,-i\partial_y)\bigr)\psi_{a\sigma}\nonumber\\&+\frac{1}{2}\sum_{a,b,c}\sum_{\sigma,\sigma'}g_{abcd}\,\bar{\psi}_{a\sigma}\bar{\psi}_{b\sigma'}\psi_{c\sigma'}\psi_{d\sigma}~.
%\end{align}
\begin{align}
\label{eqn:LE_Hamiltonian}
\hat{H}_\text{eff}&=\sum_{a}\sum_{\sigma}\int_{\mathbf{k}_a}\,\xi^{(a)}(\mathbf{k}_a)\,\hat{\psi}_{a\sigma}^\dag(\mathbf{k}_a)\hat{\psi}_{a\sigma}(\mathbf{k}_a)\nonumber\\&+\frac{1}{2}\sum_{a,b,c}\sum_{\sigma,\sigma'}\int_{\mathbf{k}_a,\mathbf{k}_b,\mathbf{k}_c}g_{abcd}\,\hat{\psi}^\dag_{a\sigma}(\mathbf{k}_a)\hat{\psi}^\dag_{b\sigma'}(\mathbf{k}_b)\times\nonumber\\&\hspace{40mm}\times\hat{\psi}_{c\sigma'}(\mathbf{k}_c)\hat{\psi}_{d\sigma}(\mathbf{k}_d)~.
\end{align}
Here, $\hat{\psi}^\dag_{a\sigma}$ and $\hat{\psi}_{a\sigma}$ are low-energy fermionic annihilation and creation operators, $\sigma$ and $\sigma'$ are spin indices, and $a$, $b$, $c$ and $d$ are patch indices. The patch indices take values in $\{1,2\}$ in the single-VHS and HOVHS cases. The momenta $\mathbf{k}_a$, $\mathbf{k}_b$, $\mathbf{k}_c$ and $\mathbf{k}_d$ are now measured relative to the centers of the relevant patches, and all integrals run over $(-k_\text{cut},k_\text{cut})\times(-k_\text{cut},k_\text{cut})$. The values of $d$ and $\mathbf{k}_d$ are set, through momentum conservation, by the values of $a$, $b$, $c$, $\mathbf{k}_a$, $\mathbf{k}_b$ and $\mathbf{k}_c$. $\xi^{(a)}$ is the dispersion in patch $a$, and the constants $g_{abcd}$ are amplitudes for inter-patch scattering. In the two-patch scenarios, symmetries dictate that there are only four independent couplings, {\it viz.\/} $g_1=g_{1111}$, $g_2=g_{1221}$, $e=g_{1212}$ and $u=g_{1122}$. The first two of these represent direct processes, $e$ corresponds to exchange, and $u$ is an umklapp coupling. See Fig.~\ref{fig:pRG_2patch_scatterings} for a pictorial representation.

We use the dimensionless RG time $y=\nu^{-1}_0\Pi_{\text{pp}}^\mathbf{0}(\Omega)=\ln^2(W/|\Omega|)$~\cite{HYFY2015,JHCHHL2016,MTCH2020}; this is zero at the beginning of the flow, and the IR limit is $y\to \infty$. The resulting flow equations are~\cite{NFTRMS1998,Irkhin_2001}
\begin{align}
\label{eqn:pRG_sing_flow}
\dot{g}_1&=-d_p^{\mathbf{0}}(g_1^2+u^2)+d_h^\mathbf{0}(g_1^2+e^2-2g_2^2+2g_2e)~,\nonumber\\
\dot{g}_2&=-d_p^{\mathbf{Q}_0}(g_2^2+e^2)+2d_h^\mathbf{0}(e-g_2)g_1+d_h^{\mathbf{Q}_0}(u^2+g_2^2)~,\nonumber\\
\dot{u}&=-2d_p^{\mathbf{0}}ug_1+2d_h^{\mathbf{Q}_0}(2g_2-e)u~,\nonumber\\
\dot{e}&=-2d_p^{\mathbf{Q}_0}g_2e+2d_h^\mathbf{0}g_1e+2d_h^{\mathbf{Q}_0}(g_2-e)e~,
\end{align}
where the dots denote $y$--derivatives. Here and throughout, we have defined the following functions of $y$:
\begin{align}
\label{eqn:d_functions}
d_p^\mathbf{q}(y)=\nu_0\,\frac{d\Pi_{\text{pp}}^\mathbf{q}(\Omega)}{d\Pi_{\text{pp}}^\mathbf{0}(\Omega)}~\,\,,~\,\,d_h^\mathbf{q}(y)=\nu_0\,\frac{d\Pi_{\text{ph}}^\mathbf{q}(\Omega)}{d\Pi_{\text{pp}}^\mathbf{0}(\Omega)}~.
\end{align}
 $d_p^{\mathbf{0}}$ identically equals $\nu_0$, while from (\ref{eqn:prg_log_susc}) we find that $d_h^\mathbf{0}\sim \nu_0/\sqrt{y}$, $d_p^{\mathbf{Q}_1}\sim \nu_0\gamma_1/\sqrt{y}$ and $d_h^{\mathbf{Q}_1}\sim \nu_0\gamma_2/\sqrt{y}$, as $y\to \infty$. At small $y$, each of the latter three functions should approach $\nu_0$~\cite{HYFY2015,JHCHHL2016} --- therefore, we use the common approximations
\begin{align}
\label{eqn:sing_interp}
d_h^\mathbf{0}= \frac{\nu_0}{\sqrt{1+y}}~\,\,~&,~\,\,~ d_p^{\mathbf{Q}_0}= \frac{\nu_0\gamma_1}{\sqrt{\gamma_1^2+y}}~,\nonumber\\ d_h^{\mathbf{Q}_0}&= \frac{\nu_0\gamma_2}{\sqrt{\gamma_2^2+y}}~,
\end{align}
which interpolate between the two limits. It is common at this point to rescale the couplings as $\nu_0\hspace{0.2mm}g_i\to g_i$, $\nu_0\hspace{0.2mm}e\to e$ and $\nu_0\hspace{0.2mm}u\to u$~\cite{NFTRMS1998,SMAC2013,XCYYHY2015,HYFY2015,YSJB2018,AZDEJB2023}; we do not do so as we wish to compare to TUFRG, which deals directly with dimensionful couplings. The initial conditions for the flows are determined by the Hubbard model (\ref{eqn:Hubbard_realspace}), so that all couplings have initial value $U$.

We solve the flow equations numerically.  The solutions for the couplings will generically diverge at some critical RG time $y_c$ (which will vary as the parameters vary). This is an artefact of our having calculated the right-hand-sides of Eq. \eqref{eqn:pRG_sing_flow} perturbatively: if they were calculated to all orders, divergences would only occur strictly in the IR limit. A common approach in pRG calculations is to study the flows of {\it ratios\/} of the couplings, since these tend to flow to constants~\cite{RNLLAC2012,ASGGCC2017,MTCH2020,ZWYWFW2023,XHASXW2023}; however, since we compare below to TUFRG we adopt the same stopping criteria for the flow as in that method. That is, we stop the flow once one of the couplings becomes greater than $1000\,t_1$ in absolute value; for the values of $U$ and the hoppings that we consider, this is a good proxy for divergence. Furthermore, if no divergence has occurred by the time the RG scale $\Omega$ has been lowered to a minimum value of $10^{-5}\,t_1$ we regard the system as still being a Fermi liquid. This, roughly, sets the temperature at $10^{-5}\,t_1/k_B$ ($\approx 120$~mK when $t_1=1~\text{eV}$, for example), since lower-energy fluctuations are not integrated out.

In order to determine which ordered phase the divergence of the couplings signifies, we must assess the divergence of the susceptibilities of the various ordering tendencies. This is effected by perturbing the Hamiltonian (\ref{eqn:LE_Hamiltonian}) by infinitesimal pairing interactions in the particle-hole and particle-particle channels and computing the one-loop flows of the gaps. For example, for spin-singlet superconductivity the perturbation is $\delta\hat{H}_{\text{eff}}=\sum_a\int_{\mathbf{k}_a} \Delta_a\hat{\psi}^\dag_{a\uparrow}(\mathbf{k})\hat{\psi}^\dag_{\bar{a}\downarrow}(-\mathbf{k})+\text{h.c.}$, where $\Delta_a$ is the gap on patch $a$ and $\bar{a}$ is the patch with opposite momentum to $a$. Computing the one-loop flows gives $\dot{\Delta}_a=\nu_0\sum_b M_{ab}(y)\Delta_b$, where the entries of the matrix $M$ are taken from the set of couplings $g_i$, $e$ and $u$. Diagonalizing, one finds two linear combinations of the gaps $\Delta_a$, each of which corresponds to a different pairing symmetry. Each also evolves independently under the RG, e.g.\ the $A_{1g}$-representation gap $\Delta_s=\sum_a\Delta_a$ satisfies $\dot{\Delta}_s=\gamma_{s\text{SC}}(y)\Delta_s$, with $\gamma_{s\text{SC}}=-\nu_0(e_1+g_2+2u)$. We call the prefactor, $\gamma_{s\text{SC}}$, the ``rate''. There is one such for every ordering channel --- they are listed in appendix~\ref{append:prg_rates}.

We take the dominant phase to be the one with the largest rate at $y=y_c$. It should be noted that these rates in fact only tell us how quickly the gap in each channel is diverging; if the gap diverges as $\Delta\sim (y_c-y)^{-\bar{\gamma}}$, then the susceptibility scales as $\chi\sim (y_c-y)^{1-2\bar{\gamma}}$. Therefore, in addition to requiring $\bar{\gamma}>0$, we also need $\bar{\gamma}>1/2$ in order for the phase to dominate over the Fermi liquid~\cite{SMAC2013,LCACCH2020,XHASXW2023}. In our approach we do not have direct access to the exponents $\bar{\gamma}$ and so cannot make this comparison --- however, the leading exponent seems to usually be greater than 1/2 in similar studies~\cite{RNLLAC2012,SMAC2013,JHCHHL2016,LCACCH2020}, so that the only likely induced error is that our scheme may estimate any Fermi-liquid regions of the phase diagram to be somewhat smaller than they ought to be.

%The framework used for the HOVHS and split-VHS cases are very similar to the above, and so we only provide the salient differences in the following.

\subsubsection{HOVHS case}

%The HOVHS case has also received attention in the literature using pRG methods~\cite{LCACCH2020,XHASXW2023}.
For the HOVHS case there are again four independent susceptibilities, of which three are divergent~\cite{LCACCH2020,XHASXW2023}:
\begin{align}
\label{eqn:prg_HOVHS_susc}
\Pi_{\text{pp}}^\mathbf{0}(\Omega)\sim \frac{C_1(N_++N_-)}{|\Omega|^{1/4}}~\,&,~\,\Pi_{\text{ph}}^\mathbf{0}(\Omega)\sim \frac{C_2(N_++N_-)}{|\Omega|^{1/4}}~,\nonumber\\\Pi_{\text{pp}}^{\mathbf{Q}_0}(\Omega)&\sim \frac{m_-^\text{X}}{2\pi}\ln\frac{W}{|\Omega|}~,
\end{align}
where $C_1\approx 2.17$, $C_2\approx 0.54$. Further,
\begin{align}
N_+=\frac{N_-}{\sqrt{2}}=\frac{\bigr[\Gamma(1/4)\bigr]^2\bigr(m_-^\text{X}\bigr)^{1/2}}{8\pi^{5/2}|t_3|^{1/4}}~,
\end{align}
where $\Gamma$ is the gamma function. We have retained all divergent terms in the above --- $\Pi_{\text{pp}}^\mathbf{0}(\Omega)$ and $\Pi_{\text{ph}}^\mathbf{0}(\Omega)$ have no logarithmically diverging parts, and $\Pi_\text{ph}^{\mathbf{Q}_0}(\Omega)$ is non-divergent. $\Pi_{\text{pp}}^{\mathbf{Q}_0}(\Omega)$ is sub-leading and so was neglected in previous studies~\cite{LCACCH2020,XHASXW2023} --- we retain it for completeness.%, and because it can affect the phase at large values of $U$~\cite{TSLRCH2026} (larger than the values considered herein).

%The HOVHS case has been studied in the literature using pRG methods~\cite{LCACCH2020,XHASXW2023} --- however, we retain all terms associated to the sub-leading $d_p^{\mathbf{Q}_0}$, which were neglected in Refs.~\cite{} but in fact significantly affect the phase diagram \textcolor{red}{safe to mention that here, as want to deal with it in other paper? If okay to mention, ``cite'' other paper (as to be published)?}. 
We now use the RG time $y=\bar{\nu}_0^{-1}[\Pi_\text{pp}^\mathbf{0}(\Omega)-\Pi_\text{pp}^\mathbf{0}(W)]$, where $\bar{\nu}_0=m_-^\text{X}/(4\pi^2)$ and $\Pi_{\text{pp}}^\mathbf{0}(\Omega)$ is now given in (\ref{eqn:prg_HOVHS_susc}). We have subtracted off the susceptibility evaluated at the high-energy cutoff $W$ in order that we have $y=0$ at the start of the flow~\cite{LCACCH2020}. There are four couplings, as for the single-VHS case, and they obey flow equations of the same form as (\ref{eqn:pRG_sing_flow}) except that $d_p^{\mathbf{0}}$ is identically $\bar{\nu}_0$, $d_h^{\mathbf{Q}_0}$ is zero and
\begin{align}
d_h^\mathbf{0}=\bar{\nu}_0\,\frac{1+C_2y/C_1}{1+y}~\,\,,\,\,~d_p^{\mathbf{Q}_0}=\frac{8\pi\bar{\nu}_0}{8\pi+y}~.
\end{align}
All of the couplings have initial value $U$ and the flow equations are solved precisely as in the single-VHS case. We also analyse the divergence of the various susceptibilities analogously --- the associated rates are given in appendix~\ref{append:prg_rates}.

\subsubsection{Split-VHS case}

In the split-VHS case there are, {\it prima facie\/}, six non-interacting susceptibilities to be calculated; however, one can show that within a patch setup $\Pi_{\text{pp}}^\mathbf{0}(\Omega)=\Pi_{\text{pp}}^{\mathbf{Q}_2}(\Omega)$ and $\Pi_{\text{ph}}^\mathbf{0}(\Omega)=\Pi_{\text{ph}}^{\mathbf{Q}_2}(\Omega)$. There are thus only four independent susceptibilities. Furthermore, they have the same forms as those in the single-VHS case, save with the effective masses at X replaced by those at $\mathcal{P}$ --- this results in $(d_h^\mathbf{0})_\text{split}=(d_h^\mathbf{0})_\text{single}$, $(d_p^{\mathbf{Q}_1})_\text{split}=(d_p^{\mathbf{Q}_0})_\text{single}$ and $(d_h^{\mathbf{Q}_1})_\text{split}=(d_h^{\mathbf{Q}_0})_\text{single}$. The $d$-functions for this case can thus be read off (\ref{eqn:gammas}) and (\ref{eqn:sing_interp}), with $\nu_0=\sqrt{m_+^\mathcal{P}m_-^\mathcal{P}}/(4\pi^2)$ and $\kappa=m_+^\mathcal{P}/m_-^\mathcal{P}$. 

\begin{figure}
\centering
\includegraphics[width=0.9\linewidth]{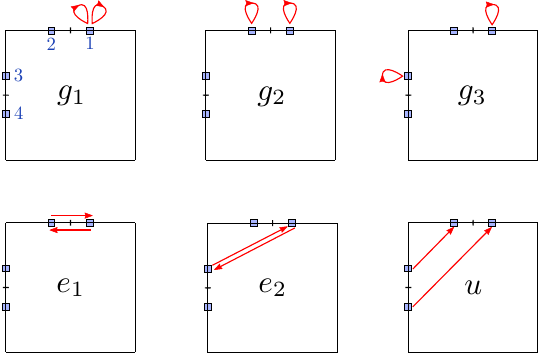}
\caption{The six inequivalent four-point couplings in the split-VHS case.}
\label{fig:pRG_4patch_scatterings}
\end{figure}

Here there are eleven independent couplings --- however, our choice of an on-site Hubbard model for the flow's initial condition dictates that a number of these couplings are exactly equal throughout. One need only consider six couplings, then, namely $g_1=g_{1111}$, $g_2=g_{1221}$, $g_3=g_{1331}$, $e_1=g_{1212}$, $e_2=g_{1313}$ and $u=g_{1234}$. The associated scattering processes are displayed schematically in Fig.~\ref{fig:pRG_4patch_scatterings}. Taking $y=\ln^2(W/|\Omega|)$, we find the following flow equations:
\begin{align}
\label{eqn:pRG_split_flow}
\dot{g}_1=&-g_1^2+d_h^{\mathbf{0}}\bigr[g_1^2+e_1^2+2e_2^2\nonumber\\&+2\left(-g_2^2-2g_3^2+g_2e_1+2g_3e_2\right)\bigr]~,\nonumber\\
\dot{g}_2=&-(e_1^2+g_2^2+2u^2)\nonumber\\&+d_h^\mathbf{0}(g_2^2-2g_1g_2+2g_1e_1-4g_3^2+4g_3e_2)~,\nonumber\\
\dot{g}_3=&-d_p^{\mathbf{Q}_1}(e_2^2+g_3^2)+d_h^{\mathbf{Q}_1}(g_3^2+u^2)\nonumber\\&+2d_h^\mathbf{0}(-g_1g_3-2g_2g_3+g_1e_2+g_2e_2+e_1g_3)~,\nonumber\\
\dot{e}_1=&-2(g_2e_1+u^2)+2d_h^{\mathbf{0}}(g_1e_1+e_2^2)\nonumber\\&+2d_h^{\mathbf{Q}_1}(g_2-e_1)e_1~,\nonumber\\
\dot{e}_2=&-2g_3e_2d_p^{\mathbf{Q}_1}+2d_h^{\mathbf{0}}(g_1e_2+e_1e_2)\nonumber\\&+2d_h^{\mathbf{Q}_1}(-e_2^2+e_2g_3)~,\nonumber\\
\dot{u}=&-2(ue_1+ug_2)+2d_h^{\mathbf{Q}_1}(2g_3u-ue_2)~.
\end{align}
See appendix~\ref{append:prg_rates} for the rates associated to the various orders.

\subsection{TUFRG}
An alternative method to predict the ordered state adopted by a system, to the same order in the renormalized couplings as in pRG, is to use the numerical TUFRG scheme \cite{Lichtenstien_TUFRG_2017,Profe_TUFRG_2022}. Full details are given in, e.g., Refs.~\cite{WMMSCH2012,Lichtenstien_TUFRG_2017,JL2018,Profe_TUFRG_2022,Beyer_Reference_2022,Profe_diverge_2024} --- however, for ease of comparison with pRG we recapitulate some of them here. One solves for $\Gamma^\Lambda(\mathbf{k}_1,\mathbf{k}_2,\mathbf{k}_3)$, the one-particle-irreducible four-point function (the analog of $g_{abcd}$ from pRG). It depends on the running RG cutoff $\Lambda$, which in our case is a frequency cutoff introduced in the action by $G(\omega,\mathbf{k})\to G^\Lambda(\omega,\mathbf{k})=G(\omega,\mathbf{k})\Theta(|\omega|-\Lambda)$. In $\Gamma^\Lambda$, $\mathbf{k}_1$ and $\mathbf{k}_2$ are the outgoing momenta, and $\mathbf{k}_3$ and $\mathbf{k}_1+\mathbf{k}_2-\mathbf{k}_3$ the incoming momenta; the first three, crucially, are allowed to take values in the whole Brillouin zone. We have neglected any frequency dependence in $\Gamma^\Lambda$, and we furthermore ignore all self-energy feedback. 

Then, one may write~\cite{Lichtenstien_TUFRG_2017,JL2018,GRHHAT2018,Hille_Quantitative_2020}
\begin{align}
&\Gamma^\Lambda(\mathbf{k}_1,\mathbf{k}_2,\mathbf{k}_3)=V^\Lambda(\mathbf{k}_1,\mathbf{k}_2,\mathbf{k}_3)+\Phi^\Lambda_P(\mathbf{k}_1+\mathbf{k}_2,\mathbf{k}_1,\mathbf{k}_4)\nonumber\\&+\Phi^\Lambda_C(\mathbf{k}_1-\mathbf{k}_3,\mathbf{k}_1,\mathbf{k}_4)+\Phi^\Lambda_D(\mathbf{k}_1-\mathbf{k}_4,\mathbf{k}_1,\mathbf{k}_3)~,
\end{align}
%Old convention
%\begin{align}
%&\Gamma^\Lambda(\mathbf{k}_1,\mathbf{k}_2,\mathbf{k}_3)=V^\Lambda(\mathbf{k}_1,\mathbf{k}_2,\mathbf{k}_3)+\Phi^\Lambda_P(\mathbf{k}_1+\mathbf{k}_2,\mathbf{k}_1,\mathbf{k}_4)\nonumber\\&+\Phi^\Lambda_C(\mathbf{k}_1-\mathbf{k}_3,\mathbf{k}_3,\mathbf{k}_4)+\Phi^\Lambda_D(\mathbf{k}_1-\mathbf{k}_4,\mathbf{k}_3,\mathbf{k}_4)~,
%\end{align}
where $\mathbf{k}_4=\mathbf{k}_1+\mathbf{k}_2-\mathbf{k}_3$. Here, $\Phi^\Lambda_i$ is the sum of all two-particle diagrams that are two-particle-reducible in channel $i$; we have $i=P$, $C$ or $D$ for the particle-particle, crossed particle-hole or direct particle-hole channels respectively (as defined in, e.g.,~\cite{GRHHAT2018,Profe_TUFRG_2022}). $V^\Lambda$ is then the sum of all two-particle-irreducible diagrams, and is approximated by the bare interaction (the Hubbard $U$ in our case). The one-loop RG equations then read~\cite{WMMSCH2012,Lichtenstien_TUFRG_2017,JL2018}
%This calculation solves the one-loop parquet equation by integrating over the Full Brillouin zone. 
\begin{align}
\label{eqn:TUFRG_floweqns}
&\partial_\Lambda \Phi^\Lambda_P(\mathbf{q}_P,\mathbf{k},\mathbf{k}')=\nonumber\\&-\int_{\bar{\mathbf{k}}}L^\Lambda_\text{pp}(\mathbf{q}_P,\bar{\mathbf{k}})\,\Gamma^\Lambda(\bar{\mathbf{k}},\mathbf{q}_P-\bar{\mathbf{k}},\mathbf{k}')\Gamma^\Lambda(\mathbf{q}_P-\mathbf{k},\mathbf{k},\mathbf{q}_P-\bar{\mathbf{k}})~,\nonumber\\
&\partial_\Lambda \Phi^\Lambda_C(\mathbf{q}_C,\mathbf{k},\mathbf{k}')=\nonumber\\&-\int_{\bar{\mathbf{k}}}L^\Lambda_\text{ph}(\mathbf{q}_C,\bar{\mathbf{k}})\,\Gamma^\Lambda(\mathbf{k}'-\mathbf{q}_C,\bar{\mathbf{k}},\mathbf{k}')\Gamma^\Lambda(\bar{\mathbf{k}}-\mathbf{q}_C,\mathbf{k},\bar{\mathbf{k}}) ~,\nonumber\\&\partial_\Lambda\Phi^\Lambda_D(\mathbf{q}_D,\mathbf{k},\mathbf{k}')=\nonumber\\&\int_{\bar{\mathbf{k}}}L^\Lambda_\text{ph}(\mathbf{q}_D,\bar{\mathbf{k}})\Bigr[2\Gamma^\Lambda(\bar{\mathbf{k}},\mathbf{k}'-\mathbf{q}_D,\mathbf{k}')\Gamma^\Lambda(\mathbf{k},\bar{\mathbf{k}}-\mathbf{q}_D,\bar{\mathbf{k}})\nonumber\\&\hspace{18mm}-\Gamma^\Lambda(\mathbf{k}'-\mathbf{q}_D,\bar{\mathbf{k}},\mathbf{k}')\Gamma^\Lambda(\mathbf{k},\bar{\mathbf{k}}-\mathbf{q}_D,\bar{\mathbf{k}})\nonumber\\&\hspace{18mm}-\Gamma^\Lambda(\bar{\mathbf{k}},\mathbf{k}'-\mathbf{q}_D,\mathbf{k}')\Gamma^\Lambda(\bar{\mathbf{k}}-\mathbf{q}_D,\mathbf{k},\bar{\mathbf{k}})\Bigr]~,
\end{align}
%Old convention
%\begin{align}
%\label{eqn:TUFRG_floweqns}
%&\partial_\Lambda \Phi^\Lambda_P(\mathbf{q}_P,\mathbf{k},\mathbf{k}')=\nonumber\\&-\int_{\bar{\mathbf{k}}}L^\Lambda_\text{pp}(\mathbf{q}_P,\bar{\mathbf{k}})\,\Gamma^\Lambda(\bar{\mathbf{k}},\mathbf{q}_P-\bar{\mathbf{k}},\mathbf{k}')\Gamma^\Lambda(\mathbf{q}_P-\mathbf{k},\mathbf{k},\mathbf{q}_P-\bar{\mathbf{k}})~,\nonumber\\
%&\partial_\Lambda \Phi^\Lambda_C(\mathbf{q}_C,\mathbf{k},\mathbf{k}')=\nonumber\\&-\int_{\bar{\mathbf{k}}}L^\Lambda_\text{ph}(\mathbf{q}_C,\bar{\mathbf{k}})\,\Gamma^\Lambda(\mathbf{k}+\mathbf{q}_C,\bar{\mathbf{k}},\mathbf{k})\Gamma^\Lambda(\bar{\mathbf{k}}+\mathbf{q}_C,\mathbf{k}'-\mathbf{q}_C,\bar{\mathbf{k}}) ~,\nonumber\\&\partial_\Lambda\Phi^\Lambda_D(\mathbf{q}_D,\mathbf{k},\mathbf{k}')=\nonumber\\&\int_{\bar{\mathbf{k}}}L^\Lambda_\text{ph}(\mathbf{q}_D,\bar{\mathbf{k}})\Bigr[2\Gamma^\Lambda(\bar{\mathbf{k}},\mathbf{q}_D+\mathbf{k}',\mathbf{k}')\Gamma^\Lambda(\bar{\mathbf{k}}+\mathbf{q}_D,\mathbf{k}-\mathbf{q}_D,\mathbf{k})\nonumber\\&\hspace{18mm}-\Gamma^\Lambda(\bar{\mathbf{k}},\mathbf{q}_D+\mathbf{k}',\mathbf{k}')\Gamma^\Lambda(\bar{\mathbf{k}}+\mathbf{q}_D,\mathbf{k}-\mathbf{q}_D,\bar{\mathbf{k}})\nonumber\\&\hspace{18mm}-\Gamma^\Lambda(\mathbf{q}_D+\mathbf{k}',\bar{\mathbf{k}},\mathbf{k}')\Gamma^\Lambda(\bar{\mathbf{k}}+\mathbf{q}_D,\mathbf{k}-\mathbf{q}_D,\mathbf{k})\Bigr]~,
%\end{align}
in which the integrals over $\bar{\mathbf{k}}$ range over the whole Brillouin zone. Furthermore,
\begin{align}
L^\Lambda_\text{pp}(\mathbf{q},\mathbf{k})&=\int_{-\infty}^\infty\frac{d\omega}{2\pi}\,\partial_\Lambda\Bigr[G^\Lambda(\omega,\mathbf{k})\,G^\Lambda(-\omega,\mathbf{q}-\mathbf{k})\Bigr]~,\nonumber\\
L_\text{ph}^\Lambda(\mathbf{q},\mathbf{k})&=\int_{-\infty}^\infty\frac{d\omega}{2\pi}\,\partial_\Lambda\Bigr[G^\Lambda(\omega,\mathbf{k})\,G^\Lambda(\omega,\mathbf{k}-\mathbf{q})\Bigr]~.
\end{align}
These loop integrals are in correspondence with (\ref{eqn:d_functions}), while (\ref{eqn:TUFRG_floweqns}) is the analogue of the pRG flow equations (\ref{eqn:pRG_sing_flow}) and (\ref{eqn:pRG_split_flow}).

The essence of the truncated-unity approach is to note that each function $\Phi_i^\Lambda(\mathbf{q},\mathbf{k},\mathbf{k}')$ varies comparatively slowly with $\mathbf{k}$ and $\mathbf{k}'$. By contrast, singularities may develop in the $\mathbf{q}$ dependence (the $\mathbf{q}$ value of the singularity is then the ordering wavevector). One may therefore~\cite{Husemann_2009,WWYXQW2012,Lichtenstien_TUFRG_2017,JL2018,Profe_TUFRG_2022,Beyer_Reference_2022} decompose the dependences on $\mathbf{k}$ and $\mathbf{k}'$ (separately) in a basis of form factors $f_n(\mathbf{k})$. Each $f_n$ corresponds to a certain neighbor distance in real space (e.g.\ $e^{\pm ik_x}$ and $e^{\pm ik_y}$ are the nearest-neighbor form factors), and so a truncation of the basis according to distance represents a controlled approximation that makes numerical integration of the flow equations feasible.

We employ TUFRG implemented within the \textsc{divERGe} software package~\cite{Profe_diverge_2024}. All calculations are performed on a two-dimensional $\mathbf{k}$-point grid ($n_k$) of $80\times80$ for the bosonic momenta $\mathbf{q}$ of the vertices, with a further refinement of $40\times40$ for the integration of the loop susceptibilities ($n_{kf}$); thus, in total, $3200\times3200$ $\mathbf{k}$-points in the Brillouin zone are used. We use a form-factor truncation of eight unit cells, which corresponds to 197 bonds per site, and define a phase transition to have occurred when either the particle-particle, particle-hole or crossed particle-hole vertex (respectively $P$, $D$ or $C$ in the notation of \textsc{divERGe}) exceeds a value of $1000\,t_1$. As for the hot-spot-pRG calculation, if no divergence has occurred by the time the flow scale has decreased below $10^{-5}\,t_1$ we declare that the phase is a Fermi liquid (FL). If a divergence does occur at a scale $\Lambda_c>10^{-5}\,t_1$, we analyse $\Gamma^{\Lambda_c}$ to deduce the ordered phase. If the divergence is in the $P$ channel we solve a linearized gap equation using the $P$-channel projection of $\Gamma^{\Lambda_c}$ to find the pairing symmetry of the dominant gap; if it is in the $D$ or $C$ channels, we construct the static charge and spin susceptibilities from the $D$-channel projection of $\Gamma^{\Lambda_c}$, and analyse their singularities~\cite{Klebl_2022,Beyer_Reference_2022,LKAFLC2023}. Further details are given in appendix~\ref{append:gaps_TUFRG}.

\section{Results}
\label{sec:Results}

\subsection{Van Hove filling}
\label{subsec:results_VHfilling}

\subsubsection{Varying $t_2/t_1$}

\begin{figure*}
\centering
\hspace{0.9mm}\includegraphics[width=0.916\linewidth]{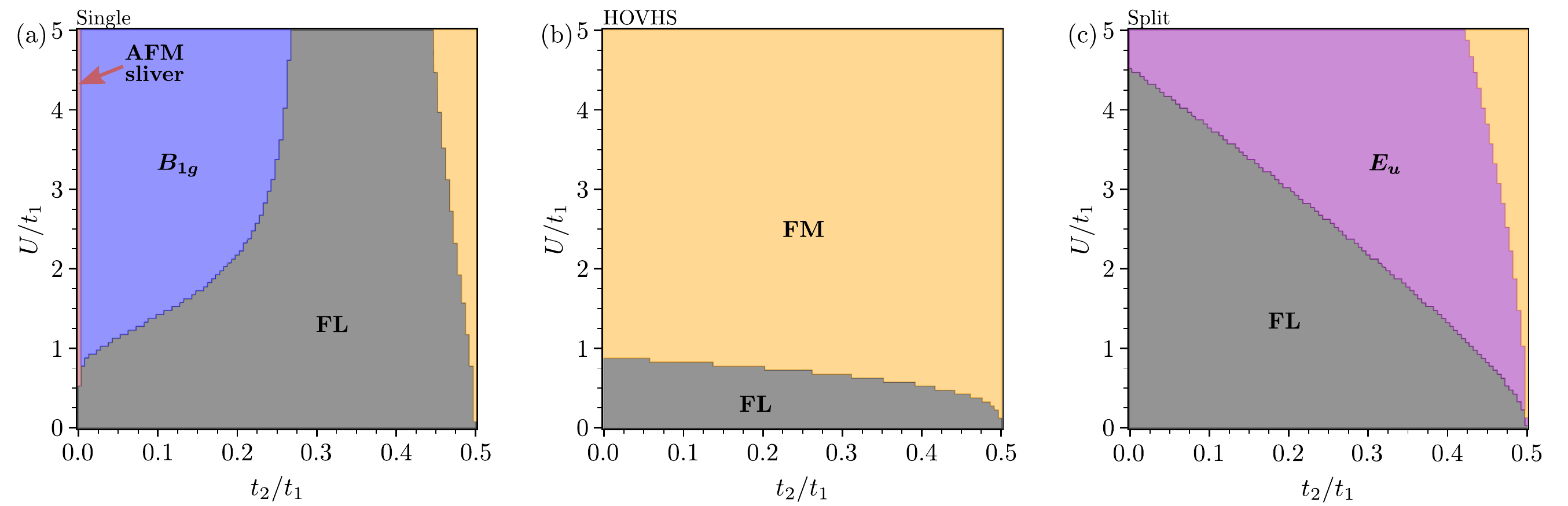}
\includegraphics[width=0.9\linewidth]{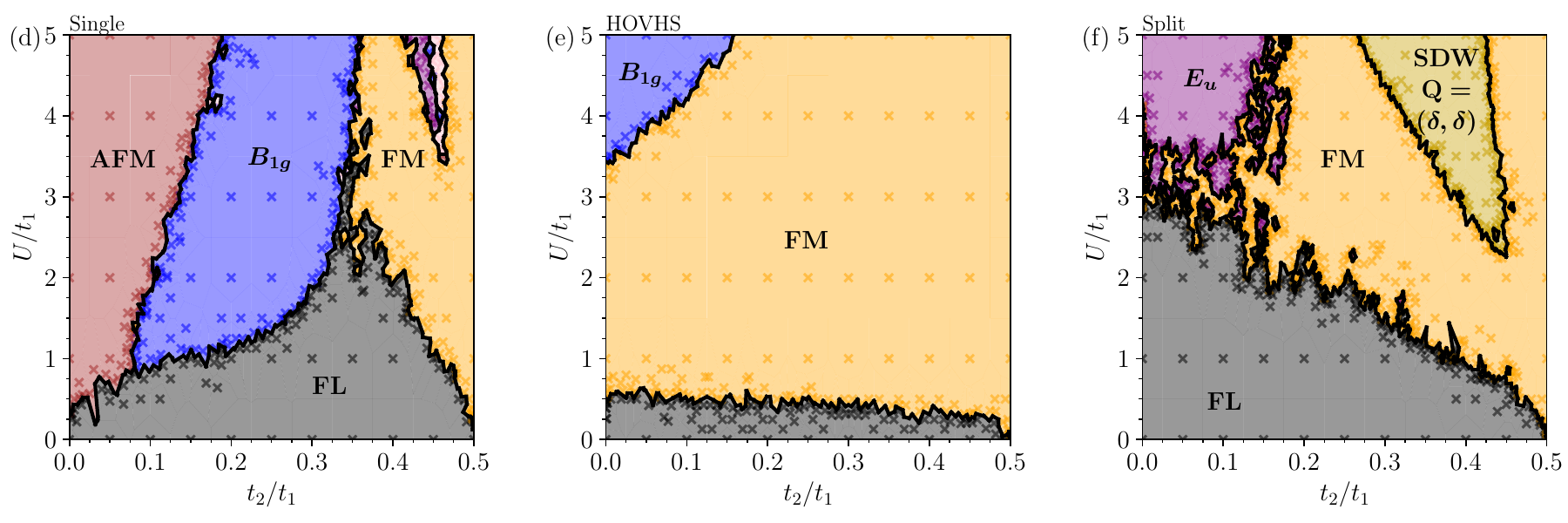}
\caption{Phase diagrams in the $(t_2/t_1,U/t_1)$ plane at Van Hove filling. Those in (a)\,--\,(c) were calculated using hot-spot pRG, for the single-VHS ($t_3=0$), HOVHS ($t_3=t_{3c}$) and split-VHS ($t_3=2t_{3c}$) cases, respectively. (d)\,--\,(f) show the equivalent phase diagrams calculated using TUFRG. The colors signify the phase: black is the Fermi-liquid state (FL), red is N{\'e}el antiferromagnetism (AFM), blue is $B_{1g}$ ($d$-wave) superconductivity, orange is ferromagnetism (FM), purple is $E_u$ ($p$-wave) superconductivity, and dark yellow is a spin-density wave (SDW) with a wavevector whose components vary but never exceed $\pi/10$ in magnitude (see appendix~\ref{append:extra_phase_diags}). The thin pink region in (d), at $t_2/t_1 \approx 0.45$ and $U\gtrsim4\,t_1$, corresponds to an $A_{1g}$ ($s$-wave) superconducting phase. For the single-VHS scenario in pRG [(a)], a thin sliver of AFM ordering occurs at $t_2/t_1 =0$.}
\label{fig:pRG_t2U}
\end{figure*}

In this section we discuss the phase diagrams in the case in which the Van Hove singularity is pinned at the Fermi level. We begin with $(t_2/t_1,U/t_1)$ phase diagrams; those calculated using pRG are displayed in Fig.~\ref{fig:pRG_t2U}(a), (b) and (c) for the single-VHS, HOVHS and split-VHS cases, respectively. In the single-VHS case (Fig.~\ref{fig:pRG_t2U}(a)), we find competition between a ferromagnetic phase and a superconducting state that transforms in the $B_{1g}$ irreducible representation of the point group $D_{4h}$ ($d$-wave). This competition appears to suppress the critical scale below $10^{-5}\,t_1$ at intermediate values of $t_2/t_1$ and has previously been reported in a number of weak-coupling RG studies of the system (e.g.\ Refs.~\cite{CHMS2001,Irkhin_2001,Husemann_2009,Giering_selfenergy_2012,Husemann_Frequency_2012,Lichtenstien_TUFRG_2017,JL2018,Beyer_Reference_2022}). 

However, this delicate interplay is completely changed in the HOVHS scenario, where now the ferromagnetic phase dominates most of phase space (Fig.~\ref{fig:pRG_t2U}(b)). This likely originates from the enhanced singularity in the DOS at the Fermi level. Intriguingly, in the split-VHS scenario (Fig.~\ref{fig:pRG_t2U}(c)), we find that the competition is now between a triplet $E_u$ superconducting state ($p$-wave) and ferromagnetism, favouring ferromagnetism only at very high values of $t_2/t_1$. We note that at $t_2/t_1=0.5$ the band structure is the same in all three cases (single-VHS, HOVHS and split-VHS), and furthermore that the dispersion is extremely flat at every point on the Fermi surface, and so the hot-spot approximation breaks down.

%The pRG calculations, however, make use of a hot-spot approximation; to test its validity, we now turn our attention to TUFRG. The fact that the latter accounts for the values of the dispersion and vertex at all points in the Brillouin zone means that it should, in principle, be more accurate, though it remains to be seen just how different their predictions are.

%This technique works in an analogous way of numerically solving the one-loop parquet equations, however it enables full-momentum zone integration of the Brillouin zone, at the loss of the analytic tractability of hot-spot pRG.
%
% This rather large change, brought about by subtle changes to the low-energy band structure around the saddle points, highlights the remarkable sensitivity of the 
%We next test the validity of the hot-spot approximation in pRG against the full momentum zone integration of TUFRG.

We next test the validity of the hot-spot method by comparing the pRG results with those of the more sophisticated TUFRG. We find broad qualitative agreement between the pRG and TURFG phase diagrams in all three scenarios shown in Fig.~\ref{fig:pRG_t2U}(d), (e) and (f), with the boundaries between the Fermi-liquid regions and correlated phases having the same general shape and in places even being quantitatively similar. This degree of general consistency between the results of the two methods is in many ways remarkable, given the simplifications entailed in the hot-spot approximation. There are two significant differences: (1) the stabilization of a sizeable N{\'e}el antiferromagnetic phase in the single-VHS scenario at low $t_2/t_1$ in the TUFRG results, with only a thin sliver of such a phase in the pRG results (Fig.~\ref{fig:pRG_t2U}(a) and (d)), and (2) the relative sizes of the triplet-$E_u$-superconductivity regions in the split-VHS case (Fig.~\ref{fig:pRG_t2U}(c) and (f)).
%the increased competition between the $E_u$-wave triplet superconducting state (labelled by the $E_u$ irreducible representation in the $D_{4h}$ point group) and the Ferromagnetic (FM) phase in the split-VHS scenario (Fig. (\ref{fig:pRG_t2U} (f)).
It appears that in the TUFRG calculation the ferromagnetic state competes more strongly with the $E_u$ superconductivity than in pRG. This is emphasized by the dendritic-looking phase boundary in the top-left of Fig.~\ref{fig:pRG_t2U}(f), at $U\gtrsim 3\,t_1$; in the pRG diagram the superconducting regime extends down to the weakest couplings.

These modifications reflect the occasional importance of full-momentum-zone integration. In the case of the first of these differences, in the single-VHS scenario the $t_2=0$ Fermi surface at Van Hove filling is a square with corners at the high-symmetry X points. Opposite sides of this square are strongly nested by the ordering wavevector $(\pi,\pi)$, leading to a stable N{\'e}el phase. This nesting is not captured within our hot-spot scheme; one could perform a separate pRG calculation using patches covering (parts of) the Fermi-surface edges~\cite{NFTR1998,SWSS2014,MTCH2021}, but it is not clear how this could be done in conjunction with the VHS patch scheme. Note that as $t_2\to 0$, $\gamma_2\to \infty$ and so $d_h^{\mathbf{Q}_0}$ becomes identically $\nu_0$, i.e. the VHS-focused calculation still captures the fact that the VHSs become nested as $t_2\to 0$. Accordingly, we do find a very thin sliver of antiferromagnetism in pRG precisely at $t_2=0$.

The second difference above is more notable, particularly because the triplet-superconductivity region predicted by pRG is quite large compared to that in the TUFRG results, and extends to very weak couplings. As stated above, close to $t_2=t_1/2$ (in all three cases) the dispersion becomes very flat across the whole Fermi surface and the hot-spot method breaks down; however, even at such moderate values as $t_2=0.375\,t_1$, pRG predicts the onset of superconductivity at $U\approx 1.55\,t_1$. This is small compared to the bandwidth (which is $8\,t_1$ at this value of $t_2/t_1$) and so firmly in the weak-coupling limit, where weak-coupling methods should be most accurate.
%The second difference above is somewhat more notable. The fact that pRG predicts a much larger triplet-superconductivity region than TUFRG is not very worrying, but its extending to such weak coupling is more so.
%weak coupling (the bandwidth is $8t_1$), and one tends to trust patch schemes most at weak coupling \textcolor{red}{citation?} --- we therefore caution that, though hot-spot pRG is a very useful method that can give good qualitative insight, TUFRG should be used whenever possible.
A possible reason for the discrepancy lies in the fact that, in the split-VHS case, there is a band maximum at each X-point. The energetic difference between this maximum and the energy of the VHSs is small compared to the overall bandwidth, and so it is plausible that the associated step in the DOS (see Fig.~\ref{fig:fig1}(c)) contributes significantly to any TUFRG flows that have critical scale $\Lambda_c/t_1 \ll 10^{-1}$. The pRG patch scheme is oblivious to this feature, however.

There is one further difference between the pRG and TUFRG results, namely that the split-VHS scenario has an appreciably sized region in which the magnetic instability acquires a finite wavevector. We find the wavevector to be of the form $\mathbf{Q}=(\delta,\delta)$, where $\delta$ is a small constant that varies across the region. $\delta$ is zero at the boundary with the ferromagnetic region and increases towards the interior of the spin-density-wave phase, but it never exceeds $\pi/10$ in absolute value. The origin of this wavevector is unclear. It could potentially arise from weak nesting across the Fermi-surface pockets around the X points in the split-VHS case (see Fig.~\ref{fig:fig1}(b)) --- however, the same phase appears in the HOVHS case away from Van Hove filling (Fig.~\ref{fig:doping}(b) below), where no such nesting occurs.
%This is similar in size to some of the smaller wavevectors spanning the Fermi-surface pockets around the X points in the split-VHS case (see Fig.~\ref{fig:fig1}(c)), suggesting that the phase may be driven by weak nesting across these pockets. 
In any case, the phase cannot be captured by the hot-spot approximation. For completeness, in appendix~\ref{append:extra_phase_diags} we plot all $(t_2/t_1,U/t_1)$ phase diagrams with $U$ extending up to $\sim125\%$ of the bandwidth.

%The weak-coupling approximation breaks down at such large $U$, and so the results there make no claim to physical relevance. In addition to the small pockets of $E_u$ and $A_{1g}$ (likely $s$-wave) superconductivity in the single-VHS case, the most salient feature is additional $B_{1g}$ superconductivity in both the HOVHS and split-VHS scenarios. In fact, we find that within this phase space there always exists a $B_{1g}$ superconducting phase followed by an antiferromagnetic phase as $U$ is increased far beyond the bandwidth (not shown).

\begin{figure}
\centering
\hspace{-0.5mm}\includegraphics[width=0.988\linewidth]{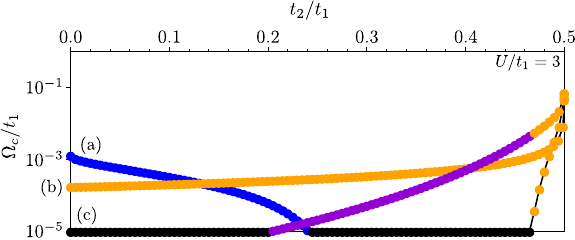}\vspace{0.5cm}
\includegraphics[width=\linewidth]{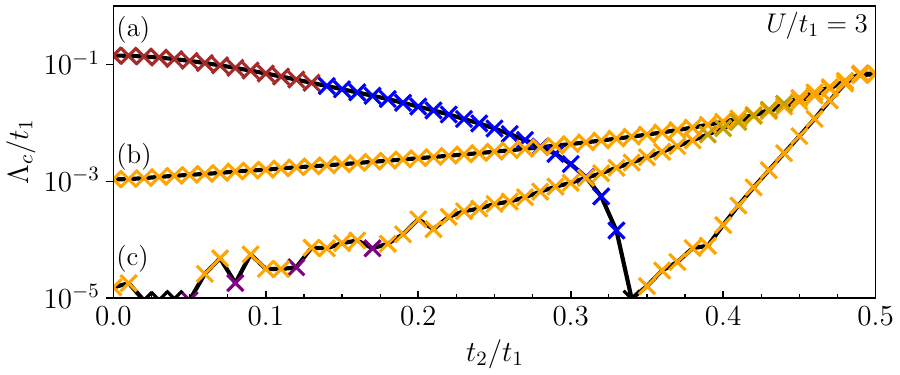}
\caption{Line cuts of the critical scale as a function of $t_2$, taken at $U=3\,t_1$, for the three cases considered in Fig.~\ref{fig:pRG_t2U}. The upper panel is the pRG result and the lower the TUFRG result, and (a), (b) and (c) are respectively the single-VHS, HOVHS and split-VHS cases.
The marker colors indicate the phase as in Fig.~\ref{fig:pRG_t2U}. Broad qualitative agreement between the two methods is apparent.}
\label{fig:prg_critscale}
\end{figure}

To further explore the agreement between the pRG and TUFRG calculations, we plot the critical scale as a function of $t_2/t_1$ for a fixed $U=3\,t_1$ in Fig.~\ref{fig:prg_critscale}. For the single-VHS scenario, this calculation has been performed using FRG in a variety of publications~\cite{CHMS2001,Husemann_2009,Giering_selfenergy_2012,Husemann_Frequency_2012,Lichtenstien_TUFRG_2017,JL2018,Beyer_Reference_2022}.

We again find qualitative agreement between the pRG and TUFRG calculations for evolution of the critical scale in all three cases; the approximate shapes of the TUFRG linecuts are roughly reproduced by pRG. That being said, there are of course definite quantitative differences --- in particular, in the single-VHS scenario the critical scale at intermediate $t_2/t_1$ is suppressed far more strongly than in TUFRG, leading to an extended Fermi-liquid regime between the $B_{1g}$ superconducting and ferromagnetic phases.

We now focus on the TUFRG calculations.
Our prediction for the evolution of the critical scale in the single-VHS case [(a)] very closely matches that presented in the literature \cite{CHMS2001,Husemann_2009,Giering_selfenergy_2012,Husemann_Frequency_2012,Lichtenstien_TUFRG_2017,JL2018,Beyer_Reference_2022}. In the HOVHS scenario we find that, despite the stabilization of ferromagnetism, the critical scale is smaller at low $t_2/t_1$ than for the antiferromagnetic or $B_{1g}$ superconducting states of the single-VHS case. The critical scale in the split-VHS scenario at low $t_2/t_1$ is even smaller still, being $\lesssim 10^{-4}\, t_1$ and changing quite discontinuously. This is a manifestation of the competition between the ferromagnetic and $E_u$ superconducting states, corresponding to the dendritic phase boundaries in Fig.~\ref{fig:pRG_t2U}(f). Higher $\mathbf{k}$-space resolution might smoothen these phase boundaries --- however, even if this were so, the fact that more than $3200\times3200$ $\mathbf{k}$-points would be required for the loop integrals is itself a sign that there is intrinsically strong competition there.

\subsubsection{Varying $t_3/t_1$}

\begin{figure}
\centering
\hspace{0.8mm}\includegraphics[width=0.939\linewidth]{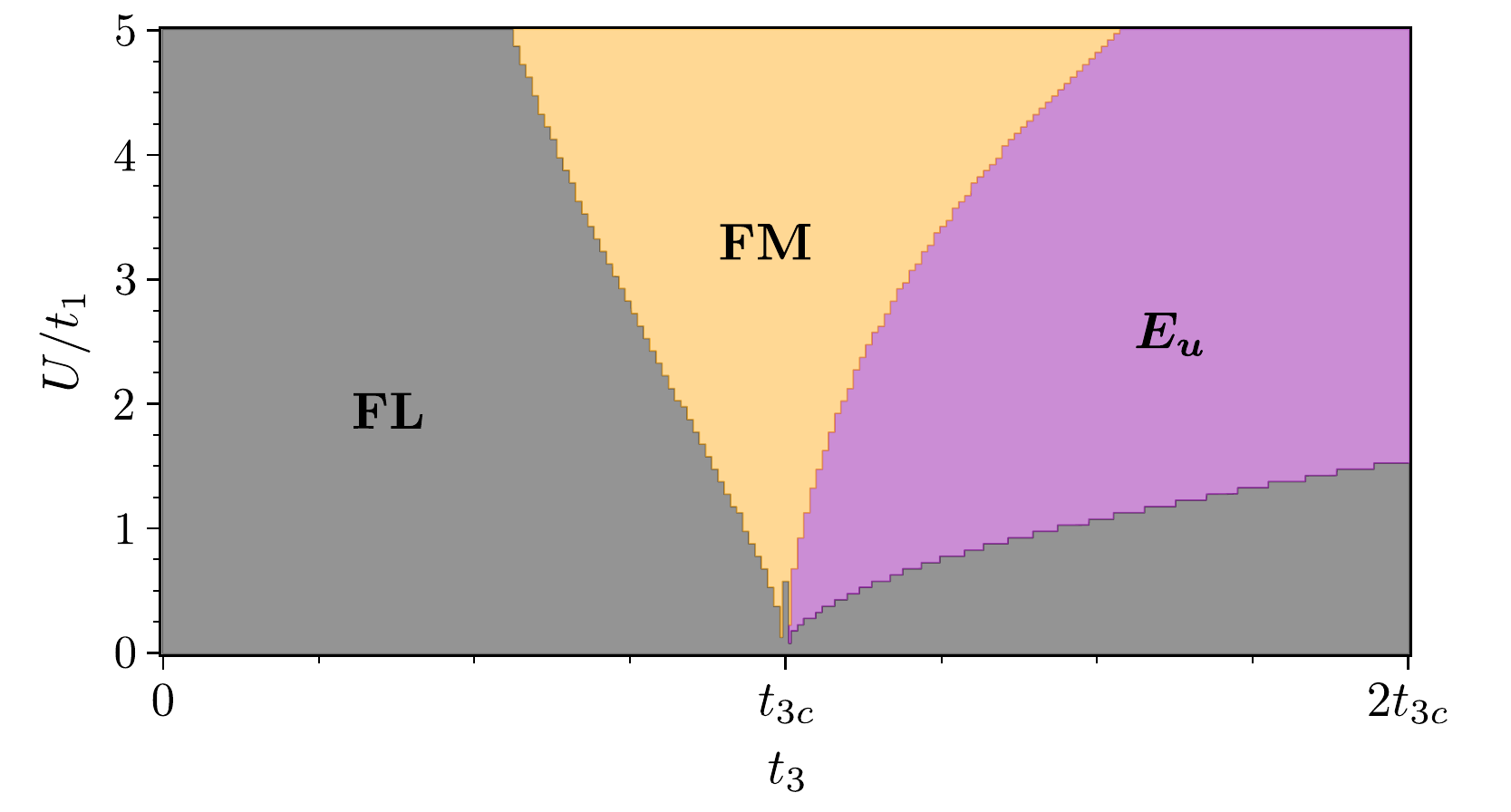}
\includegraphics[width=0.869\linewidth]{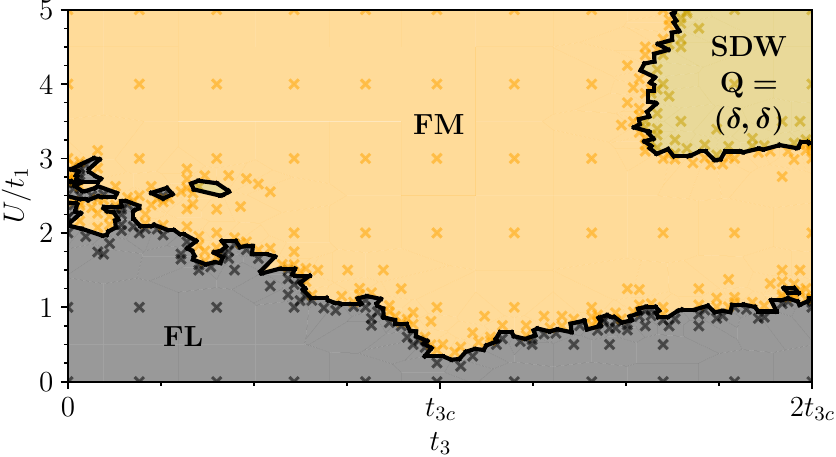}
\caption{The $(t_3/t_1,U/t_1)$ phase diagram, as predicted by pRG (upper panel) and TUFRG (lower), with $t_2/t_1=0.375$. The labeling and coloring of phases is as in Fig.~\ref{fig:pRG_t2U}. Note the discontinuity in the boundary of the Fermi-liquid regime in pRG at $t_3=t_{3c}$ --- see the main text for details.}
\label{fig:pRG_t3U_fig}
\end{figure}

Thus far, we have ``discretely'' switched between the single-VHS, HOVHS and split-VHS cases. However, one might wish to study how the ground state changes as one continuously tunes $t_3/t_1$ to move between the three cases, at fixed $t_2/t_1$. Granted, in the previous section the value of $t_3/t_1$ changed when we changed $t_2/t_1$ in the HOVHS and split-VHS cases --- however, this never corresponded to continuously changing the VHSs from ordinary to higher-order. There are many viable values to which we could fix $t_2/t_1$; in the interest of brevity we choose only one, namely $t_2/t_1=0.375$. This value was chosen as it ensures $t_1>2t_2>t_3$ for all values of $t_3$ considered, and is approximately the ratio used in realistic tight binding models for e.g.\ the $d_{xy}$ band of \ce{Sr2RuO4}~\cite{TSJRSS2014}. The $(t_3/t_1,U/t_1)$ phase diagrams at this value of $t_2/t_1$ are displayed in Fig.~\ref{fig:pRG_t3U_fig}. Note that, in the pRG figure, we have only used the results of the HOVHS-case calculations precisely at $t_3=t_{3c}$, using those of the single-VHS or split-VHS cases everywhere else (even though those analyses become quite ill-motivated close to the HOVHS case). This results in a slight discontinuity in the edge of the Fermi-liquid region at $t_3=t_{3c}$.

We again mostly find good qualitative agreement between pRG and TUFRG --- the shapes (though not the positions) of the boundaries of the Fermi-liquid regions are similar, and both methods exhibit a large ``fan'' of ferromagnetism at intermediate $t_3/t_1$, with its boundary being suppressed to near $U=0$ in the HOVHS limit ($t_3=t_{3c}$). We again see an appreciable $E_u$ superconducting regime in the pRG results and not those of TUFRG --- as before, this raises some concerns as to how much one can trust pRG, particularly because there is no triplet region anywhere in the TUFRG results, at this value of $t_2/t_1$. Our earlier comments on the possible origin of this discrepancy still apply. It is important to note that, as $t_3$ approaches $t_{3c}$ from above, the patch approximation of the split-VHS case becomes increasingly poor, as the split VHSs lying on either side of each X point become ever flatter and move closer together. This is particularly relevant in the triplet region. Consider, say, the patches at $(\pm k_\mathcal{P},\pi)$: because the superconducting gap is odd, it must have opposite signs on either of these patches. Therefore, a node must occur between them --- however, this configuration is likely to cost an increasingly large amount of free energy as the patches move closer together, a fact to which the pRG is totally insensitive. Note in particular that triplet superconductivity is impossible in pRG when the patches are at time-reversal invariant points like X~\cite{HYFY2015}.

The TUFRG phase diagram provides insight into the sensitivity of the ground state to the type and positions of the system's VHSs. The type clearly matters, as exemplified by Fig.~\ref{fig:pRG_t2U} --- however, the TUFRG diagram in Fig.~\ref{fig:pRG_t3U_fig} demonstrates that the dependence of the phase on $t_3$ is mostly quite smooth, with all displayed regions having appreciable extent along the $t_3$ direction. In particular, there are almost no discontinuous changes in the diagram at $t_3=t_{3c}$, i.e.\ apart from the strong suppression of the boundary of the Fermi-liquid region, the effects one might have ascribed to the HOVHS seem to be realized in a sizeable range of nearby $t_3$ values. Of course, this might be particular to our chosen value of $t_2/t_1$ --- we leave it to future work to conduct a more thorough analysis.

\subsection{Away from Van Hove filling}
\label{subsec:doping}

\begin{figure*}
\centering
\includegraphics[width=0.9\linewidth]{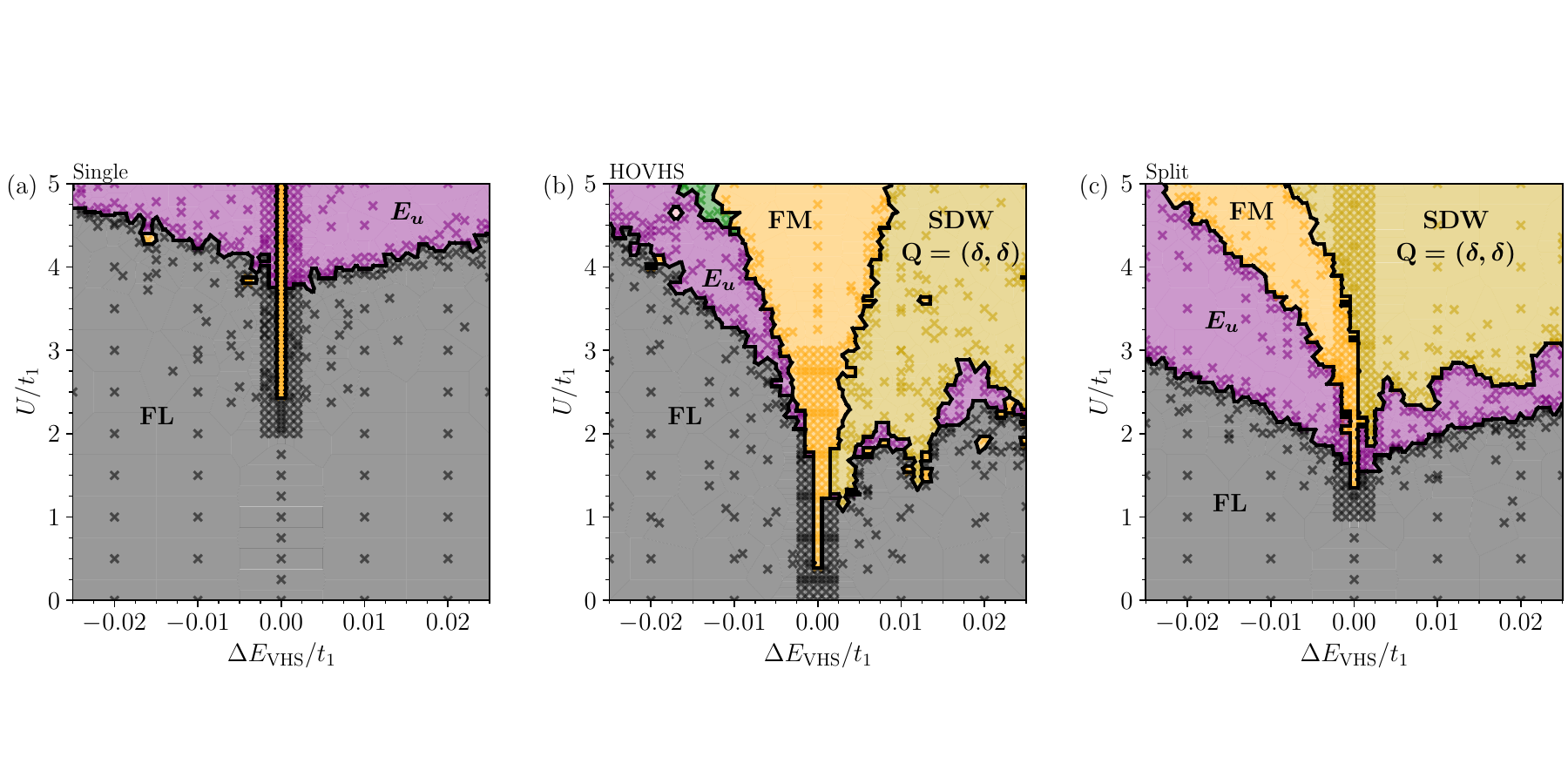}
\caption{Phase diagrams in the $(\Delta E_\text{VHS}/t_1,U/t_1)$ plane. $\Delta E_\text{VHS}$ is an energy shift added to the dispersion; for $\Delta E_\text{VHS}>0$ the VHSs lie above the Fermi level, and vice versa. (a) is the single-VHS case ($t_3=0)$, (b) the HOVHS case ($t_3=t_{3c}$) and (c) the split-VHS case ($t_3=2t_{3c}$). Here, we fix $t_2/t_1=0.375$, and the labeling of phases is as in Fig.~\ref{fig:pRG_t2U}. Green signifies a charge-density wave phase whose wavevector varies across the displayed region. The wavevector in fact becomes $\mathbf{0}$ in a few places, mostly near $U/t_1=4.5$ --- further classification is required to decide what type of Pomeranchuk instability this is (see appendix~\ref{append:gaps_TUFRG}), but since this does not affect our conclusions we do no more analysis.}
\label{fig:doping}
\end{figure*}

A setup in which a VHS is pinned to the Fermi level acts as a useful idealization in which to assess the maximum possible effect the VHS could in principle have ---
%and has been studied extensively e.g in Refs \cite{Honerkamp_Temperature_2001,Husemann_2009,Giering_selfenergy_2012,Husemann_Frequency_2012,Beyer_Reference_2022} 
however, in a real material it is nigh impossible to tune a VHS precisely to $\varepsilon_F$. This raises the question of how robust the phase diagrams in Figs.~\ref{fig:pRG_t2U} and \ref{fig:pRG_t3U_fig} are with respect to minor doping perturbations. To this end, in Fig.~\ref{fig:doping} we show the doping-vs-$U$ phase diagrams, using the fixed ratio $t_2/t_1=0.375$ as earlier. The doping change has been effected by adding an energetic shift $\Delta E_\text{VHS}$ to the non-interacting dispersion, so that when $\Delta E_{\text{VHS}}>0$ the VHSs lie above the Fermi level. Although it is possible to study the doping dependence using pRG \cite{YSJB2018}, the hot-spot approximation becomes less accurate the further the Van Hove singularities are detuned from the Fermi level. We therefore present henceforth only results from the full-band-structure TUFRG calculations.   

We find that the lowest-$U$ ordered state --- which is always ferromagnetic when the VHS is at $\varepsilon_F$ --- is very sensitive to small amounts of detuning from Van Hove filling. This is dramatically apparent in the single-VHS case, Fig.~\ref{fig:doping}(a). We find that the ferromagnetic ground state is suppressed even for shifts as small at $0.001\,t_1$, which, if we assume $t_1=1~\text{eV}$, means a shift in the chemical potential of only $1~\text{meV}$. This strong sensitivity arises because the DOS singularity is logarithmic, decaying more rapidly with energy than, e.g.\ a HOVHS. Thus, the large DOS required to stabilize ferromagnetic order only occurs in a very narrow energy range about the VHS in the single-VHS case.

Indeed, though the ferromagnetic state is still very sensitive to doping for $U\lesssim 2\,t_1$ in the HOVHS scenario (Fig.~\ref{fig:doping}(b)), at larger values of $U$ there is a more appreciable region of phase space that achieves ferromagnetism. The spin-density wave that occurs at positive $\Delta E_\text{VHS}$ is also related to this, exhibiting a wavevector of $\mathbf{Q}=(\delta,\delta)$ that is non-zero but small:\ $\delta$ is zero at the boundary with the ferromagnetic region, only increasing towards the interior of the spin-density-wave phase. %The spin-density wave that occurs at positive $\Delta E_\text{VHS}$ is also related to this, exhibiting a small but finite wavevector of $\mathbf{Q}=(\delta,\delta)$ (where $\delta \leqslant\pi/10$) due to the nesting with the saddle point away from the high-symmetry point. 
A similarly sized ferromagnetic region is seen in the split-VHS case, Fig.~\ref{fig:doping}(c) --- this might appear to be at odds with our above statements, since the VHSs in this case are also only logarithmic. However, the band maxima at the X point (see Fig.~\ref{fig:fig1}(c)) may play a role here. At small $\Delta E_\text{VHS}<0$ the Fermi level lies energetically between the VHSs and the maxima; at sufficiently large $U$ the system may not be sensitive to the comparatively small energetic difference between these features and so the areas around the X and $\mathcal{P}$ points may act as  flat-band regions at $\varepsilon_F$, enhancing ferromagnetism. This would accord with the observation that the ferromagnetic region in Fig.~\ref{fig:fig1}(c) moves to more negative $\Delta E_\text{VHS}$ as $U$ increases. 

At any rate, the ferromagnetic phase is still quite sensitive to detuning from Van Hove filling in all three cases. On the other hand, many of the other phases --- those occurring away from Van Hove filling and the non-ferromagnetic phases at larger $U$ (see Fig.~\ref{fig:doping_largeU} in appendix~\ref{append:extra_phase_diags}) --- seem to be in general somewhat more robust. We particularly note the prevalence of the triplet superconducting state of $E_u$ ($p$-wave) symmetry, arising from ferromagnetic fluctuations associated to the nearby magnetic regions.%This phase has been seen in the literature in the single-VHS case~\cite{}.

%At larger $U$ (Fig.~\ref{fig:doping_largeU}(a)) there is also an appreciable $B_{2g}$ ($d$-wave) superconducting state and an incommensurate SDW phase (with $\mathbf{Q}=(0.245\pi,\pi)$).

The behavior of the phase diagrams in Fig.~\ref{fig:doping} under a sign change of $\Delta E_\text{VHS}$ is also interesting. The DOS in the single-VHS scenario is mostly symmetric about its logarithmic divergence; this results in a doping phase diagram that is left-right symmetric for small $\Delta E_\text{VHS}$. However, as is well known~\cite{NYHILF2019,HILF2019,ACASJB2020,NYLF2020,XHASXW2023} (see also Fig.~\ref{fig:fig1}(c)), the diverging part of the DOS around a HOVHS is not symmetric in energy. We find that this produces an asymmetry in the phase diagram in Fig.~\ref{fig:doping}(b); it matters here whether the VHS is above or below the Fermi level. For example, when it is below we tend to obtain an $E_u$ superconducting state, while when it is above we have a small-$\mathbf{Q}$ spin-density wave. In the split-VHS scenario the leading part of the DOS is again symmetric around the divergence --- however, the nearby maxima at positive
energies (see Fig.~\ref{fig:fig1}(c)) do contribute a significant asymmetry. This is a weaker asymmetry than in the HOVHS case and only becomes important when $U$ is large enough; accordingly, the phase diagram Fig.~\ref{fig:doping}(c) has a weaker asymmetry than the HOVHS case that only sets in at $U\gtrsim2\,t_1$.

\section{Discussion}
\label{sec:discussion}

Significant research has been conducted into how the ground state of a material evolves as a quasi-two-dimensional VHS is pushed through the Fermi level, particularly in such materials as \ce{Sr2RuO4}~\cite{Steppke_Strong_2017,MBAGYM2018,Chandrasekaran_Engineering_2024}, \ce{Sr3Ru2O7}~\cite{KISSTH2007,HYAK2007,Efremov_Multicritical_2019} and kagome systems~\cite{HYZHYZ2022,Profe_Kagome_2024,DKHNYS2025}. The calculations presented here show that the phase realized is quite sensitive to detuning of a VHS from the Fermi level; this resonates with the common experimental experience that the ground states of quantum materials are very sensitive to doping. While a good deal of sensitivity persists in the HOVHS and split-VHS cases, we suggest that engineering such scenarios might help reduce the amount of fine tuning needed to realize specific phases. The fact that the phase diagram is quite asymmetric around Van Hove filling in the HOVHS case (and to some extent in the split-VHS case) may be a useful insight for material design once extended to fully realistic electronic structures. Furthermore, it may offer a qualitative explanation as to why \ce{Sr3Ru2O7} (which has a VHS below the Fermi level) exhibits paramagnetic behavior, whereas \ce{Sr2RuO4} (which has a VHS above the Fermi level) displays superconducting behavior. 

Controlling material properties through manipulation of HOVHSs has become a topic of particular importance: it was suggested that HOVHSs may be relevant to superconductivity in the ruthenates, influence the ground state of doped graphene-based systems, and be used to control topology in twisted bilayer kagome-based materials~\cite{Efremov_Multicritical_2019,LCACCH2020,Classen_HighOrder_2025,Perkins_Designing_2025}. Since bringing about HOVHSs at the Fermi level requires more fine tuning of the material's electronic structure than for ordinary VHSs, one could question how experimentally realistic the possibility is --- however, evidence of materials featuring HOVHSs has begun to emerge~\cite{Classen_HighOrder_2025}. The fact that, in the doping-vs-$U$ phase diagrams (Fig.~\ref{fig:doping}), the more stable, non-ferromagnetic phases are still definitely influenced by the presence of the VHSs (which determines, e.g.\ the (a)symmetry under $E_\text{VHS}\to -E_\text{VHS}$) suggests that even if a HOVHS exists in a material's band structure but is not precisely at the Fermi level, it may be possible to use its presence to manipulate the ground state if the interactions are sufficiently strong. The instability of the ferromagnetic state upon small amounts of doping (and so a reduction in the DOS at the Fermi level) may also be related to avoidance of the ferromagnetic Stoner instability at single Van Hove points~\cite{Ojajarvi2024,tupitsyn2026}.

In our model, the bandwidth is always $8\,t_1$ (except in the split-VHS case, where it occasionally increases (to no more than $9\,t_1$)). There is considerable uncertainty as to how strong interactions can be before the predictions of one-loop (weak-coupling) RG for finite-density fermions can no longer be trusted~\cite{MSCH2001,Hille_Quantitative_2020,AGNREW2024}. We generically have confidence in the method up to values of $U$ of roughly half the bandwidth, but at larger $U$ values results are likely to be significantly modified by higher-loop terms neglected here. However, in appendix~\ref{append:extra_phase_diags} we present the full phase diagrams up to $\sim 125\,\%$ of the bandwidth, for the sake of completeness and for the interested reader.

The pRG implemented within a hot-spot framework has a long history of application to fermionic lattice models~\cite{HS1987,PLGMDP1987,NFTR1998,NFTRMS1998,Irkhin_2001,KLTR2009,RNLLAC2012,SMAC2013,SWSS2014,XCYYHY2015,HYFY2015,JHCHHL2016,ASGGCC2017,YSJB2018,WQLLZZ2019,MTCH2020,LCACCH2020,MTCH2021,ZWYWFW2023,XHASXW2023,AZDEJB2023,Lee_Unified_2025}. This approximation was chiefly used in the interest of simplicity, or because full momentum resolution was computationally unfeasible; all the same, it has undoubtedly made invaluable contributions to the understanding of many materials and phenomena. Here we have shown that, for systems with well-defined hot spots, the approximation is mostly fairly accurate in broad qualitative terms, but under certain circumstances its neglect of most of the band structure can qualitatively alter the predicted phase diagram. Of course, we have drawn this conclusion based on a single model --- however, the square-lattice Hubbard model with third-nearest-neighbor hopping is by no means exotic, and the near-Fermi-level energetic landscapes present here (e.g.\ split VHSs with nearby maxima) are surely pertinent to other systems. The advent of TUFRG, with its full-momentum-zone integration, allows us to move beyond hot-spot schemes.%, and our results suggest that whenever possible, TUFRG should be used when implementing weak-coupling finite-density fermionic RG. 

%-----------------------
%acknowledgement
%-----------------------

\begin{acknowledgments}
We thank Joseph Betouras, Matthew Bunney, David Perkins and Peter Wahl for useful discussions. LCR and TPS acknowledge EPSRC funding through UKRI3345. TPS additionally received EPSRC funding under grant EP/T518062/1. Part of this work was carried out on the Hypatia computing cluster of the University of St Andrews.
\end{acknowledgments}

%-----------------------
%Appendices
%-----------------------

\begin{appendices}
\makeatletter\renewcommand*{\@seccntformat}[1]{
\MakeUppercase{\appendixname} \thesection:
}\makeatother

%%%%%%%%%% Prefix a A,B, etc to all equations, figures, tables and reset the counters %%%%%%%%%%
\makeatletter
\renewcommand{\theequation}{\thesection\arabic{equation}}
\renewcommand{\thefigure}{\thesection\arabic{figure}}
%Removed this as need to be able to place floats in appendix to which they don't belong, for better placement of actual figures.
\renewcommand{\bibnumfmt}[1]{[\thesection#1]}

\setcounter{equation}{0}
\setcounter{figure}{0}
\setcounter{table}{0}
\section{Derivation of non-interacting susceptibilities}
\label{append:prg_correct_suscs}

Here, we derive the leading singular parts of the inter-patch susceptibilities $\Pi_\text{ph}^{\mathbf{Q}_0}(\Omega)$ and $\Pi_\text{pp}^{\mathbf{Q}_0}(\Omega)$ occurring in the single-VHS case in section~\ref{subsec:prg} (these respectively equal, \emph{mutatis mutandis}, the functions $\Pi_\text{ph}^{\mathbf{Q}_1}(\Omega)$ and $\Pi_\text{pp}^{\mathbf{Q}_1}(\Omega)$ occurring in the split-VHS case). We also provide some commentary on the sub-leading part of $\Pi_{\text{pp}}^{\mathbf{0}}(\Omega)$.

\begin{figure*}
\centering
\subcaptionbox{\label{fig:prg_regionsa}}{\includegraphics[width=0.4\textwidth]{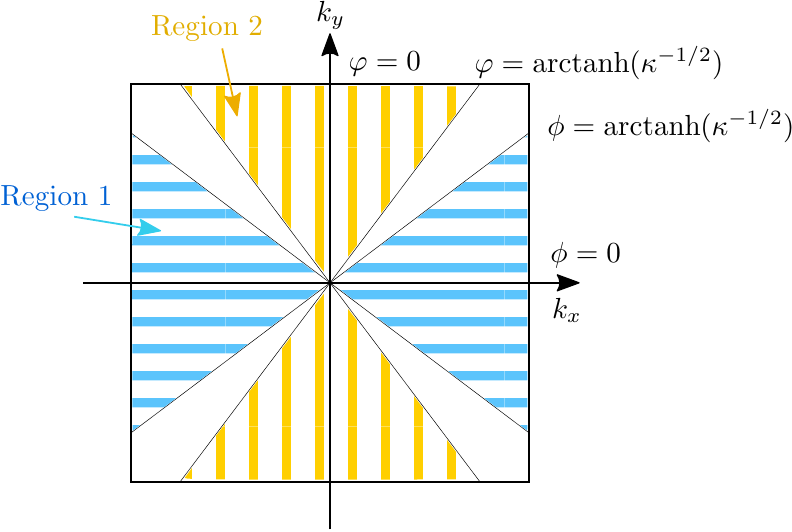}}
\hspace{5mm}\subcaptionbox{\label{fig:prg_regionsb}}{\includegraphics[width=0.43\textwidth]{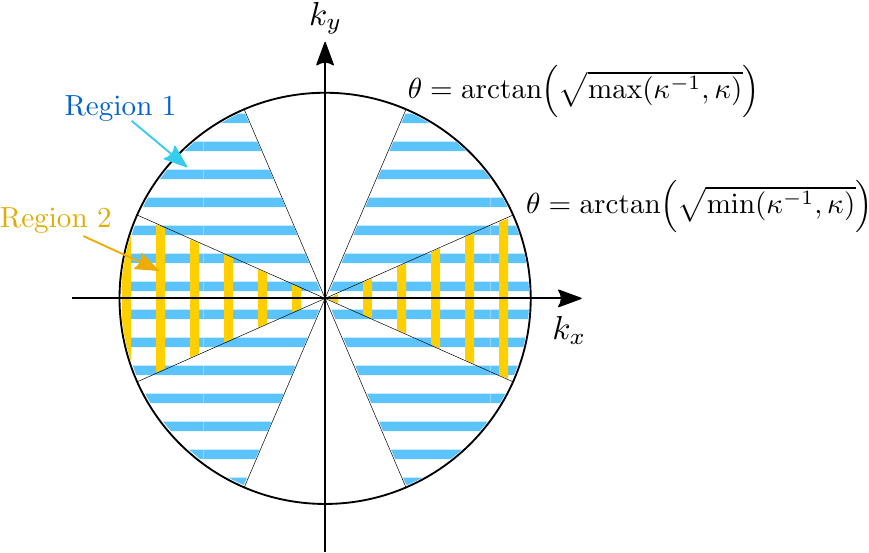}}
\caption{The integration regions and supports of the $\Theta$-functions featuring in the calculation of (a) $\Pi_{\text{ph}}^{\mathbf{Q}_0}(\Omega)$ and (b) $\Pi_{\text{pp}}^{\mathbf{Q}_0}(\Omega)$.}
\end{figure*}

We begin with $\Pi_{\text{ph}}^{\mathbf{Q}_0}(\Omega)$. We have, at zero temperature,
\begin{align}
\Pi_{\text{ph}}^{\mathbf{Q}_0}(\Omega)=\int_\mathbf{k}\frac{\Theta\left(\frac{k_x^2}{2m_+}-\frac{k_y^2}{2m_-}\right)-\Theta\left(-\frac{k_x^2}{2m_-}+\frac{k_y^2}{2m_+}\right)}{-i\Omega+\frac{1}{2}(m_+^{-1}+m_-^{-1})(k_x^2-k_y^2)}~,
\end{align}
where $\mathbf{k}$ is now a momentum-space coordinate referenced to the center of one of the patches, and $k_x,k_y\in (-k_\text{cut},k_\text{cut})$. We have also dropped superscripts `X' on the effective masses for clarity. Then, let us first assume that $\kappa=m_+/m_->1$, so that the first $\Theta$-function above is only non-zero for $|k_x|>\kappa^{1/2}|k_y|$, and the second for $|k_y|>\kappa^{1/2}|k_x|$. The associated regions in $\mathbf{k}$-space (regions 1 and 2, respectively) are non-overlapping, as shown in Fig.~\ref{fig:prg_regionsa}. Evenness of the integrand further allows us to reduce the integration to being taken over the upper right quadrant. We next introduce hyperbolic coordinates, with $(k_x,k_y)=r\,(\cosh\phi,\sinh\phi)$ on the intersection of region 1 with the upper right quadrant and $(k_x,k_y)=\rho\,(\sinh\varphi,\cosh\varphi)$ on the same for region 2. $\phi$ and $\varphi$ range over $(0,\text{arctanh}(\kappa^{-1/2}))$, while we take $r$ and $\rho$ to range over $(0,k_\text{cut})$. Strictly, the upper limit on $r$ and $\rho$ should have some dependence on the angle-like variables, but neglecting it does not change the leading (low-energy) singularity. Then, we find
\begin{align}
&\Pi_{\text{ph}}^{\mathbf{Q}_0}(\Omega)\approx \frac{m_+^{-1}+m_-^{-1}}{2\pi^2}\,\text{arctanh}(\kappa^{-1/2})\,\times\nonumber\\&\times\int_0^{k_\text{cut}^2}dx~\frac{x}{\Omega^2+x^2(m_+^{-1}+m_-^{-1})^2/4}\nonumber\\&\sim 2\nu_0\gamma_2\ln\left(\frac{W}{|\Omega|}\right)
\end{align}
to leading order, with $\nu_0$ and $\gamma_2$ given in the main text. The calculation proceeds along similar lines for $\kappa<1$.

Next, we have
\begin{align}
\Pi_\text{pp}^{\mathbf{Q}_0}(\Omega)=\int_\mathbf{k}\frac{\Theta\left(\frac{k_x^2}{2m_-}-\frac{k_y^2}{2m_+}\right)-\Theta\left(\frac{k_x^2}{2m_+}-\frac{k_y^2}{2m_-}\right)}{i\Omega+\frac{1}{2}(m_-^{-1}-m_+^{-1})(k_x^2+k_y^2)}~.
\end{align}
Here, we introduce circular polar coordinates $(k_x,k_y)=r\,(\cos\theta,\sin\theta)$ for the integration region, and approximate the latter by a circle of radius $k_\text{cut}$ (again, this does not affect the leading singular part). The $\Theta$-functions in this case restrict the integral to the symmetric difference of their supports, as shown in Fig.~\ref{fig:prg_regionsb} (which region is the support of which $\Theta$-function depends on whether $\kappa>1$ or $\kappa<1$). Again restricting to the upper right quadrant, we get
\begin{align}
\Pi_{\text{pp}}^{\mathbf{Q}_0}(\Omega)\approx& ~4\int_0^{k_\text{cut}}\frac{dr\,r}{2\pi}\int_{\arctan a}^{\arctan b}\frac{d\theta}{2\pi}\frac{\text{sgn}(\kappa-1)}{i\Omega+\frac{1}{2}(m_-^{-1}-m_+^{-1})r^2}\nonumber\\=&~\frac{(\arctan b-\arctan a)\hspace{0.5mm}\text{sgn}(\kappa-1)}{\pi^2~m_+^{-1}(\kappa-1)}\nonumber~\times\\&\times\Biggr[\frac{1}{2}\ln\Biggr(1+\Biggr(\frac{k_\text{cut}^2(m_-^{-1}-m_+^{-1})}{2\Omega}\Biggr)^2\,\Biggr)\nonumber\\&-i\arctan\left(\frac{k_\text{cut}^2(m_-^{-1}-m_+^{-1})}{2\Omega}\right)\Biggr]~,
\end{align}
where $a=[\min(\kappa,\kappa^{-1})]^{1/2}$, $b=[\max(\kappa,\kappa^{-1})]^{1/2}$ and sgn is the signum function. The leading part of this is $2\nu_0\gamma_1\ln(W/|\Omega|)$.

We conclude this appendix with a brief discussion of the sub-leading, logarithmic part of $\Pi_\text{pp}^{\mathbf{0}}(\Omega)$ in the split-VHS and single-VHS cases, as well as the high-energy cutoff $W$ in the main text. As mentioned in the main text, one should ideally keep the logarithmic part of $\Pi_\text{pp}^\mathbf{0}(\Omega)$ in the flow equations, since it is of the same order as other retained terms. However, its being sub-leading means that it depends sensitively on high-energy features of the band structure. Let us attempt a calculation of it. Using the square-patch regularization, the DOS per patch is, at small $\omega>0$, 
\begin{align}
N(\omega)&=\int_{\mathbf{k}}~\delta\left(\omega+\frac{k_y^2}{2m_-}-\frac{k_x^2}{2m_+}\right)\nonumber\\&=4\nu_0\ln\left(\sqrt{\frac{k_{\text{min}}^2}{2m_+\omega}-1}+\frac{k_{\text{min}}}{\sqrt{2m_+\omega}}\right)~,
\end{align}
with $k_{\text{min}}=\text{min}(k_\text{cut},[\kappa(k_\text{cut}^2+2m_-\omega)]^{1/2})$, and the effective masses corresponding either to an X point or $\mathcal{P}$. For $\kappa>1$, as is true in this paper, one finds
\begin{align}
\label{eqn:N(omega)}N(\omega)= 2\nu_0\ln\frac{k_\text{cut}^2}{2m_+\omega}+4\nu_0\ln 2+\mathcal{O}(\omega)~,
\end{align}
leading to $4\nu_0(\ln 2)\ln(W/|\Omega|)$ for the logarithmic part of $\Pi_\text{pp}^\mathbf{0}(\Omega)$ (adopting the convention $\Pi_\text{pp}^\mathbf{0}(\Omega)=-\int_{-W}^Wd\omega\,N(\omega)\,\text{sgn}(\omega)/(i\Omega-2\omega)$). One finds the same result if one assumes $\omega<0$ above, i.e. the constant part of $N(\omega)$ is well defined.

However, note that the half-bandwidth $W$ of each patch is not well defined, being taken here as of the order of $k_\text{cut}^2/(2m_+)$ or $k_\text{cut}^2/(2m_-)$ (ignoring the fact that the former goes to zero as we approach the HOVHS case). In (\ref{eqn:N(omega)}) we took $W=k_\text{cut}^2/(2m_+)$, but different choices are possible. Suppose we have two candidate definitions $W_1$ and $W_2$ for the half-bandwidth. Then,
\begin{align}
\ln^2\frac{W_1}{|\Omega|}=\ln^2\frac{W_2}{|\Omega|}+2\ln\biggr[\frac{W_1}{W_2}\biggr]\ln\frac{W_2}{|\Omega|}+\ln^2\frac{W_1}{W_2}~,
\end{align}
i.e.\ the coefficient of the leading, log-squared term is unchanged by switching definition, but that of the logarithmic part is not. %This illustrates why one does not need a rigorous definition of the half-bandwidth for a leading pRG calculation, but does when considering sub-leading terms. 
However, we can at least say that the logarithmic part of $\Pi_\text{pp}^\mathbf{0}(\Omega)$ equals $\nu_0 C\ln(W/|\Omega|)$, for $C$ dimensionless and of either sign, for which $|C|$ is usually of order one. Changing the half-bandwidth changes $C$ by $2\ln(W_1/W_2)$, which will be of order one for most likely choices of $W_1$ and $W_2$ (except near the HOVHS case).

Let us assume this form and assess the extent to which changing $C$ could change the results. We should now set $y=\ln^2(W/|\Omega|)+C\ln(W/|\Omega|)$, which is not monotonic in $|\Omega|$ for $C<0$, so let us further assume that taking $C>0$ captures the most important effects of including the logarithmic term. Then, the new asymptotic $d$-functions are $d_h^\mathbf{0}\sim 2\nu_0/\sqrt{C^2+4y}$, $d_p^{\mathbf{Q}_0}\sim 2\nu_0\gamma_1/\sqrt{C^2+4y}$ and $d_h^{\mathbf{Q}_0}\sim 2\nu_0\gamma_2/\sqrt{C^2+4y}$ as $y\to \infty$ (for the single-VHS case, say). However, if we are in the regime in which these asymptotic forms are valid, then $y$ must be much larger than one, and so it is not internally consistent to retain $C^2$ when compared to $4y$. In fact, taking $C$ to be zero can be seen to induce a similar level of approximation as is entailed in using the interpolating forms (\ref{eqn:sing_interp}). Since we make this level of approximation in our calculation, we may as well take $C=0$, i.e.\ neglect the sub-leading part of $\Pi_\text{pp}^\mathbf{0}(\Omega)$.

\setcounter{equation}{0}
\setcounter{figure}{0}
\setcounter{table}{0}
\section{``Rates'' in pRG}
\label{append:prg_rates}

Here we list the rates (as defined in the main text) of destabilization to the model's various ordered phases. All couplings and $d$-functions are understood to be evaluated at $y=y_c$, the RG time at which a coupling first exceeds $1000\,t_1$ in absolute value. For the single-VHS and HOVHS cases we have the following:
\begin{align*}
\begin{array}{cc}
\text{Order} & \text{Rate} \\
\hline
\Tstrut
A_{1g}\text{ SC} & -d_p^\mathbf{0}(g_1+u)\\
\Tstrut B_{1g}\text{ SC} &-d_p^\mathbf{0}(g_1-u)\\
\Tstrut A_{1g}\text{ POM} &-d_h^\mathbf{0}(g_1+2g_2-e)\\
\Tstrut B_{1g}\text{ POM} &-d_h^\mathbf{0}(g_1-2g_2+e)\\
\Tstrut \text{FM} &d_h^\mathbf{0}(g_1+e)\\
%\Tstrut B_{1g}\text{ FM} &d_h^\mathbf{0}(g_1-e)\\
\Tstrut \mathbf{Q}_0\text{-CDW} &-d_h^{\mathbf{Q}_0}(2e-g_2+ u)\\
\Tstrut \mathbf{Q}_0\text{-SDW} & d_h^{\mathbf{Q}_0}(g_2+ u)\\
\Tstrut \mathbf{Q}_0\text{-FFLO} & -d_p^{\mathbf{Q}_0}(g_2+ e)\\
\end{array}
\begin{array}{c}
\\
\\
\\
\\
\\
\\
\\
\\
\\
\\
\\
~.
\end{array}
\end{align*}
%\begin{align}
%\begin{array}{cc}
%\text{Order} & \text{Rate} \\
%\hline
%\Tstrut
%A_{1g}\text{ SC} & -d_p^\mathbf{0}(g_1+u)\\
%\Tstrut B_{1g}\text{ SC} &-d_p^\mathbf{0}(g_1-u)\\
%\Tstrut A_{1g}\text{ POM} &-d_h^\mathbf{0}(g_1+2g_2-e)\\
%\Tstrut B_{1g}\text{ POM} &-d_h^\mathbf{0}(g_1-2g_2+e)\\
%\Tstrut \text{FM} &d_h^\mathbf{0}(g_1+e)\\
%\Tstrut B_{1g}\text{ FM} &d_h^\mathbf{0}(g_1-e)\\
%\Tstrut (\mathbf{Q}_0)_{\pm}\text{-CDW} &-d_h^{\mathbf{Q}_0}(2e-g_2\pm u)\\
%\Tstrut (\mathbf{Q}_0)_{\pm}\text{-SDW} & d_h^{\mathbf{Q}_0}(g_2\pm u)\\
%\Tstrut (\mathbf{Q}_0)_{\pm}\text{-FFLO} & -d_p^{\mathbf{Q}_0}(g_2\pm e)\\
%\end{array}
%\begin{array}{c}
%\\
%\\
%\\
%\\
%\\
%\\
%\\
%\\
%\\
%\\
%\\
%\\
%~.
%\end{array}
%\end{align}
Here, ``POM'' signifies a Pomeranchuk instability in the density channel. The simplest function transforming under $B_{1g}$ is a $d$-wave form factor, and so the $B_{1g}$ POM state may correspond to an electron nematic --- however, we cannot be sure, since pRG does not provide us with the behavior of the order parameter away from the patches. The $A_{1g}$ POM state corresponds to a uniform shift in the quasiparticle density on both patches; this has at least two possible interpretations. One possibility is that the full order-parameter function becomes negative away from the patches, in such a way that the total change in density is zero, in line with Luttinger's theorem. Alternatively, the full order parameter may indeed be constant across the entire Brillouin zone --- such a state would violate Luttinger's theorem, but could be a proxy for a charge-density wave with wavevector very close to $\mathbf{Q}=\mathbf{0}$, i.e.\ a phase-separated state. 

Note that $d_h^{\mathbf{Q}_0}\equiv 0$ in the HOVHS case, and so charge-density waves (CDWs) and spin-density waves (SDWs) cannot occur there. In addition, here and below we refer to the finite-momentum particle-particle instabilities as ``FFLO'' --- whether the state realized is of the FF or LO kinds, or some other form, should be decided by more sophisticated analysis. Finally, it should be mentioned that various potential ordered states have been neglected in the above list --- for example, a $B_{1g}$ ferromagnetic state, or finite-$\mathbf{Q}$ instabilities with non-$A_{1g}$ form factors. We neglect these as we presume that, for the most part, such exotic states should not occur for weak positive couplings with such simple momentum dependence as the Hubbard interaction.

The rates in the split-VHS are as follows:
\begin{align*}
\begin{array}{cc}
\text{Order} & \text{Rate} \\
\hline
\Tstrut
A_{1g}\text{ SC} &-d_p^\mathbf{0}(e_1+g_2+2u) \\
\Tstrut B_{1g}\text{ SC} &-d_p^\mathbf{0}(e_1+g_2-2u) \\
\Tstrut E_u\text{ SC} &d_p^\mathbf{0}(e_1-g_2) \\
\Tstrut A_{1g}\text{ POM} &-d_h^\mathbf{0}(-e_1-2e_2+g_1+2g_2+4g_3) \\
\Tstrut B_{1g}\text{ POM} &-d_h^\mathbf{0}(-e_1+2e_2+g_1+2g_2-4g_3) \\
\Tstrut E_u\text{ POM} &-d_h^\mathbf{0}(e_1+g_1-2g_2) \\
\Tstrut \text{FM} &-d_h^\mathbf{0}(-e_1-2e_2-g_1) \\
\Tstrut\mathbf{Q}_1\text{-CDW} &-d_h^{\mathbf{Q}_1}(2e_2-g_3+ u)\\
\Tstrut\mathbf{Q}_1\text{-SDW} & -d_h^{\mathbf{Q}_1}(-g_3- u)\\
\Tstrut \mathbf{Q}_2\text{-CDW}&-d_h^{\mathbf{0}}(2e_1-g_2)\\
\Tstrut\mathbf{Q}_2\text{-SDW}&d_h^{\mathbf{0}}g_2\\
\Tstrut2\mathbf{K}_1\text{-FFLO} &-g_1\\
\Tstrut\mathbf{Q}_1\text{-FFLO} & -d_p^{\mathbf{Q}_1}(g_3+ e_2)\\
\end{array}
\begin{array}{c}
\\
\\
\\
\\
\\
\\
\\
\\
\\
\\
\\
\\
\\
\\
\\
\\
\\
\\
~,
\end{array}
\end{align*}
%\begin{align}
%\begin{array}{cc}
%\text{Order} & \text{Rate} \\
%\hline
%\Tstrut
%A_{1g}\text{ SC} &-d_p^\mathbf{0}(e_1+g_2+2u) \\
%\Tstrut B_{1g}\text{ SC} &-d_p^\mathbf{0}(e_1+g_2-2u) \\
%\Tstrut E_u\text{ SC} &d_p^\mathbf{0}(e_1-g_2) \\
%\Tstrut A_{1g}\text{ POM} &-d_h^\mathbf{0}(-e_1-2e_2+g_1+2g_2+4g_3) \\
%\Tstrut B_{1g}\text{ POM} &-d_h^\mathbf{0}(-e_1+2e_2+g_1+2g_2-4g_3) \\
%\Tstrut E_u\text{ POM} &-d_h^\mathbf{0}(e_1+g_1-2g_2) \\
%\Tstrut \text{FM} &-d_h^\mathbf{0}(-e_1-2e_2-g_1) \\
%\Tstrut B_{1g}\text{ FM} &-d_h^\mathbf{0}(-e_1+2e_2-g_1) \\
%\Tstrut E_u\text{ FM} & -d_h^\mathbf{0}(e_1-g_1)\\
%\Tstrut(\mathbf{Q}_1/\mathbf{Q}_2)_{\pm}\text{-CDW} &-d_h^{\mathbf{Q}_1}(2e_2-g_3\pm u)\\
%\Tstrut(\mathbf{Q}_3)_{\pm}\text{-CDW} & -d_h^{\mathbf{Q}_1}(2e_2-g_3\pm u)  \\
%\Tstrut(\mathbf{Q}_1/\mathbf{Q}_2)_{\pm}\text{-SDW} & -d_h^{\mathbf{Q}_1}(-g_3\mp u)\\
%\Tstrut(\mathbf{Q}_3)_{\pm}\text{-SDW} &-d_h^{\mathbf{Q}_1}(-g_3\mp u)\\
%\Tstrut \mathbf{Q}_4\text{-CDW}&-d_h^{\mathbf{0}}(2e_1-g_2)\\
%\Tstrut\mathbf{Q}_4\text{-SDW}&d_h^{\mathbf{0}}g_2\\
%\Tstrut2\mathbf{K}_1\text{-FFLO} &-g_1\\
%\Tstrut(\mathbf{Q}_1)_{\pm},(\mathbf{Q}_2)_{\pm},&\multirow{2}{*}{$-d_p^{\mathbf{Q}_1}(g_3\pm e_2)$}\\
%(\mathbf{Q}_3)_{\pm}\text{-FFLO} & \\
%\end{array}
%\begin{array}{c}
%\\
%\\
%\\
%\\
%\\
%\\
%\\
%\\
%\\
%\\
%\\
%\\
%\\
%\\
%\\
%\\
%\\
%\\
%\\
%\\
%\\
%\\
%\\
%\\
%~,
%\end{array}
%\end{align}
where again we have neglected the $B_{1g}$ ferromagnetic phase and the more exotic $\mathbf{Q}\neq\mathbf{0}$ states. Note that $E_u$ is a two-dimensional irreducible representation (spanned by, e.g., $p_x$ and $p_y$), and the rates for the two independent orders (in both the SC and POM cases) are degenerate under pRG. An analysis by Landau theory might lift the degeneracy~\cite{RNLLAC2012,HYFY2015,JHCHHL2016}, but that is beyond our scope. The $2\mathbf{K}_1$-FFLO phase above is degenerate with the $2\mathbf{K}_a$-FFLO phases for $a=2$, $3$ and $4$.

%OLD FLOAT ENVIRONMENTS FOR FIGURES IN APPENDIX D.
%RESTORE IF MY WORKAROUND DOESN'T WORK
%
%--------------------------------------------%
%These figures belong to appendix D, but I've put them here to improve float placement

%\begin{figure*}[t]
%\centering
%\hspace{-0mm}\includegraphics[width=0.915\textwidth]{Figures/t2U_correctOmin_phasediag_final_nested_Latex.pdf}
%\caption{$t_2$ vs $U$ phase diagrams from pRG, with $U$ ranging up to $\sim 125\,\%$ of the bandwidth. These are in correspondence with Fig.~\ref{fig:pRG_t2U}(a)\,--\,(c) of the main text.}
%\label{fig:pRG_t2vsU_largeU}
%\end{figure*}

%\begin{figure*}[t]
%\centering
%\includegraphics[width=0.9\linewidth]{Figures/t1t2_PhaseDiagram_figure_v3.pdf}
%\caption{$t_2$ vs $U$ phase diagrams from TUFRG with $U$ ranging up to $\sim 125\,\%$ of the bandwidth, corresponding to Fig.~\ref{fig:pRG_t2U}(d)\,--\,(f). In (a) the SDW region at $t_2/t_1\approx0.375$ and $U\gtrsim 8~\text{eV}$ has wavevector $\mathbf{Q}=(0.245\pi,\pi)$, while the nearby pink region is an $A_{1g}$ ($s$-wave) superconducting state. The SDW regime in (c) has a small wavevector whose components never exceed $\pi/10$ in absolute value.}
%\label{fig:TUFRG_t2vsU_largeU}
%\end{figure*}

%-------------------------------------------%

\setcounter{equation}{0}
\setcounter{figure}{0}
\setcounter{table}{0}
\section{Determination of superconducting gap symmetries and particle-hole-channel wavevectors in TUFRG}
\label{append:gaps_TUFRG}

In a TUFRG flow, if the vertex $\Gamma^{\Lambda_c}$ has diverged at some scale $\Lambda_c>10^{-5}\,t_1$ we determine which phase the model adopts by examining channel projections of $\Gamma^{\Lambda_c}$. This analysis is implemented within the divERGe library~\cite{Profe_diverge_2024}.

If the divergence is in the $P$ channel (at $\mathbf{Q}=\mathbf{0}$), we identify the symmetry of the superconducting order parameter by solving the linearized gap equation for the particle-particle pairing vertex $\Gamma^{\Lambda_c}(\mathbf{k},-\mathbf{k},-\mathbf{k}')$ at the critical cutoff scale~\cite{Beyer_Reference_2022,Klebl_2022}. Note that TUFRG is also capable of detecting finite-$\mathbf{Q}$ particle-particle instabilities --- however, none were found in our calculations. More specifically, a singular-value decomposition of the matrix product of the vertex with a particle-particle loop yields its left singular vectors; the $\mathbf{k}$-space symmetry of the eigenvector with the leading eigenvalue is the pairing symmetry. Within the TUFRG framework, the gap equation is first solved in real space, where matrix and vector sizes are set by the form-factor cutoff (in this paper, this corresponds to a real-space distance of 8 times the lattice constant). We then Fourier-transform the obtained real-space eigenvectors into momentum space; from there we characterize the symmetry of the superconducting order parameter according to the irreducible representations of the $D_{4h}$ point group. 

If the divergence occurs in the $C$ or $D$ channels we calculate static susceptibilities from the direct particle-hole vertex, $\Gamma^{\Lambda_c}(\mathbf{k}+\mathbf{Q},\mathbf{k}'-\mathbf{Q},\mathbf{k}')$~\cite{GGGV2008,Beyer_Reference_2022,LKAFLC2023}.
%The magnetic susceptibilites are obtained by considering the particle-hole vertex and calculating the four-point susceptibility at the end of the flow,
The dominant peak in the charge/spin susceptibility then corresponds to the wavevector of the charge-/spin-density wave: for example, in the magnetic sector, we have a N{\'e}el antiferromagnet if the wavevector is $(\pi,\pi)$, ferromagnetism if it is $(0,0)$, and a generic spin-density wave for any other wavevector in the Brillouin zone. Note that the charge and spin susceptibilities are mostly only sensitive to instabilities with $s$-wave form factors; one should in principle also compute, for example, the nematic charge susceptibility~\cite{DMAC2010} to detect $d$-wave Pomeranchuk instabilities. However, except for the few points mentioned in the caption of Fig.~\ref{fig:doping}, the direct particle-hole vertex at $\mathbf{Q}=\mathbf{0}$ never diverged in our calculations, so Pomeranchuk instabilities mostly never occurred.

\setcounter{equation}{0}
\setcounter{figure}{0}
\setcounter{table}{0}
\section{Phase diagrams over larger variable ranges}
\label{append:extra_phase_diags}

%OLD FLOAT ENVIRONMENTS FOR FIGURES IN APPENDIX D.
%RESTORE IF MY WORKAROUND DOESN'T WORK
%
%\begin{figure*}
%\centering
%\includegraphics[width=0.9\linewidth]{Figures/t1t2_doping_figure_150meV.pdf}
%\caption{Doping dependence of the ordered state over a larger range of doping and $U$, in correspondence with Fig.~\ref{fig:doping}. Light blue here is a $B_{2g}$ superconducting state. Note that, in the SDW regions in (b) and (c), $\delta$ attains larger values than in Fig.~\ref{fig:doping}, reaching $\approx \pi/5$.}
%\label{fig:doping_largeU}
%\end{figure*}

In this appendix we extend some of the phase diagrams in the main text to larger axis ranges. In particular, all diagrams here range up to $U=10\,t_1$, which is slightly larger than the bandwidth. As per our comments in the main text, the results of pRG and TUFRG cannot be trusted at such large $U$; we present them here solely for the interested reader. The labeling of phases and coloring of the associated regions of phase space are explained in the captions of Figs.~\ref{fig:pRG_t2U}, \ref{fig:pRG_t2vsU_largeU}, \ref{fig:TUFRG_t2vsU_largeU} and \ref{fig:doping_largeU}.

In Figs.~\ref{fig:pRG_t2vsU_largeU} and \ref{fig:TUFRG_t2vsU_largeU} we display the $(t_2/t_1,U/t_1)$ phase diagrams. One can see that the disagreement between pRG and TUFRG becomes more pronounced at large $U$; in particular, pRG misses many of the superconducting phases seen by TUFRG at $U\gtrsim 6\,t_1$ (in all three cases). It is also noteworthy that the antiferromagnetic region in the pRG results remains extremely thin; it only begins to extend to $t_2>0$ at larger $U$ still. However, since these
\onecolumngrid
\begin{center}

\hspace{-0mm}\includegraphics[width=0.915\textwidth]{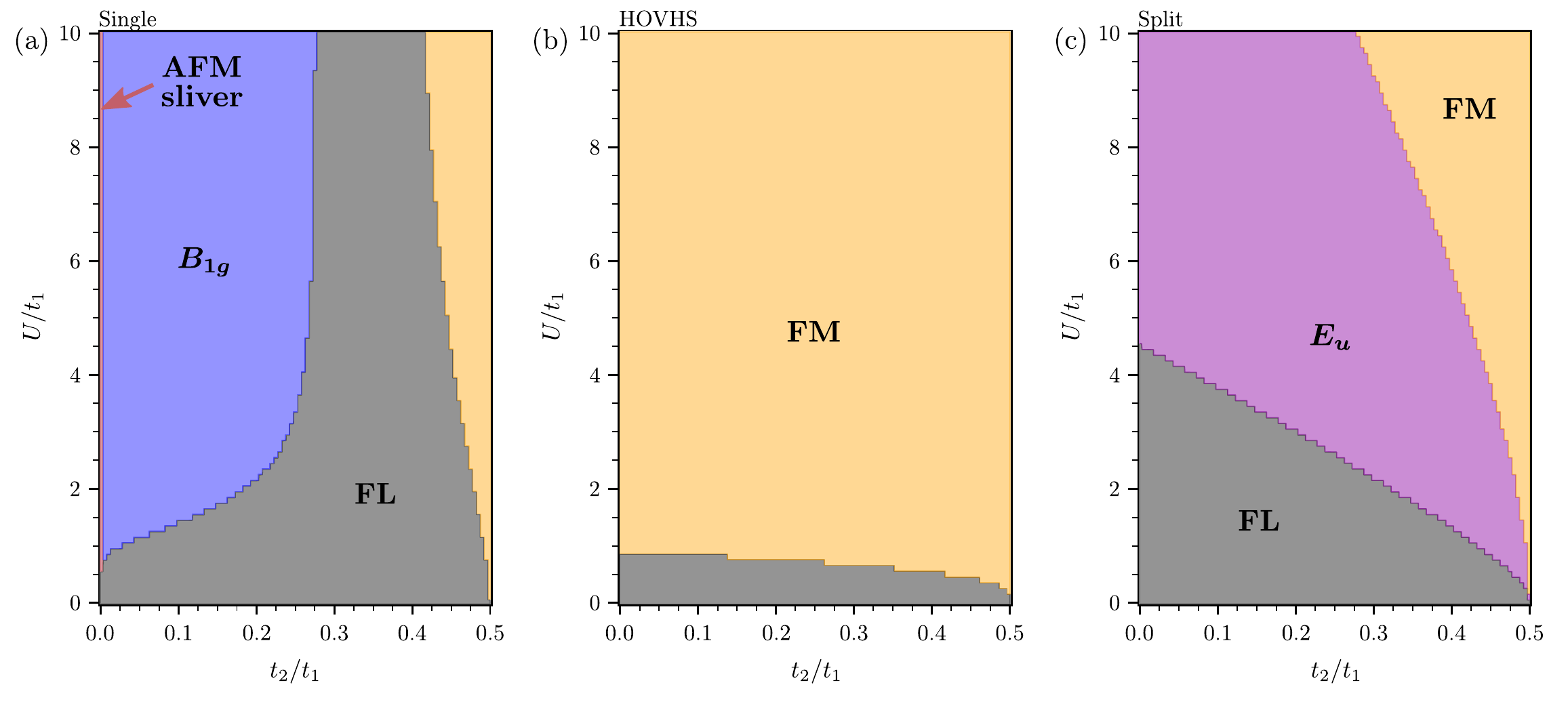}
\captionof{figure}{\centering$t_2$ vs $U$ phase diagrams from pRG, with $U$ ranging up to $\sim 125\,\%$ of the bandwidth. These are in correspondence with Fig.~\ref{fig:pRG_t2U}(a)\,--\,(c) of the main text.}
\label{fig:pRG_t2vsU_largeU}

\includegraphics[width=0.9\linewidth]{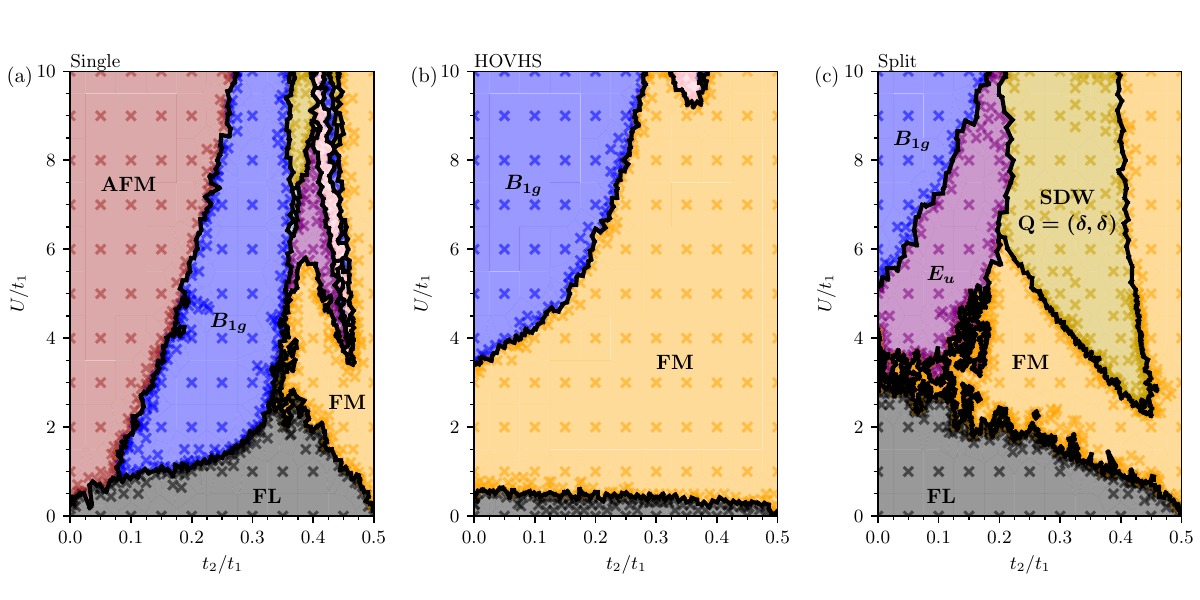}
\captionof{figure}{\centering$t_2$ vs $U$ phase diagrams from TUFRG with $U$ ranging up to $\sim 125\,\%$ of the bandwidth, corresponding to Fig.~\ref{fig:pRG_t2U}(d)\,--\,(f). In (a) the SDW region at $t_2/t_1\approx0.375$ and $U\gtrsim 8~\text{eV}$ has wavevector $\mathbf{Q}=(0.245\pi,\pi)$, while the nearby pink region is an $A_{1g}$ ($s$-wave) superconducting state. The SDW regime in (c) has a small wavevector whose components never exceed $\pi/10$ in absolute value.}
\label{fig:TUFRG_t2vsU_largeU}

\end{center}
\twocolumngrid
\noindent features are outside the domain of validity of both pRG and TUFRG they cannot contribute to any discussion on which method is more accurate. In Fig.~\ref{fig:TUFRG_t2vsU_largeU}(c) the wavevector of the displayed spin-density-wave state varies with $t_2/t_1$ and $U/t_1$, but always points along a lattice diagonal and so can be taken to be of the form $(\delta,\delta)$. $\delta$ is zero on the phase boundary with the ferromagnetic region, and increases with separation from this boundary. 

In Fig.~\ref{fig:doping_largeU} we present the doping-dependence phase diagrams over a larger range of $\Delta E_\text{VHS}/t_1$. In Fig.~\ref{fig:doping_largeU}(b) and (c), we obtain a spin-density-wave region with wavevector $(\delta,\delta)$. $\delta$ is again zero at the ferromagnet-phase boundary and increases with separation therefrom; as a result, $\delta$ is almost a monotonically increasing function of $\Delta E_\text{VHS}/t_1$, with maximum value $\approx \pi/5$ (in the displayed range of $\Delta E_\text{VHS}/t_1$). 

\onecolumngrid
\begin{center}
\includegraphics[width=0.9\linewidth]{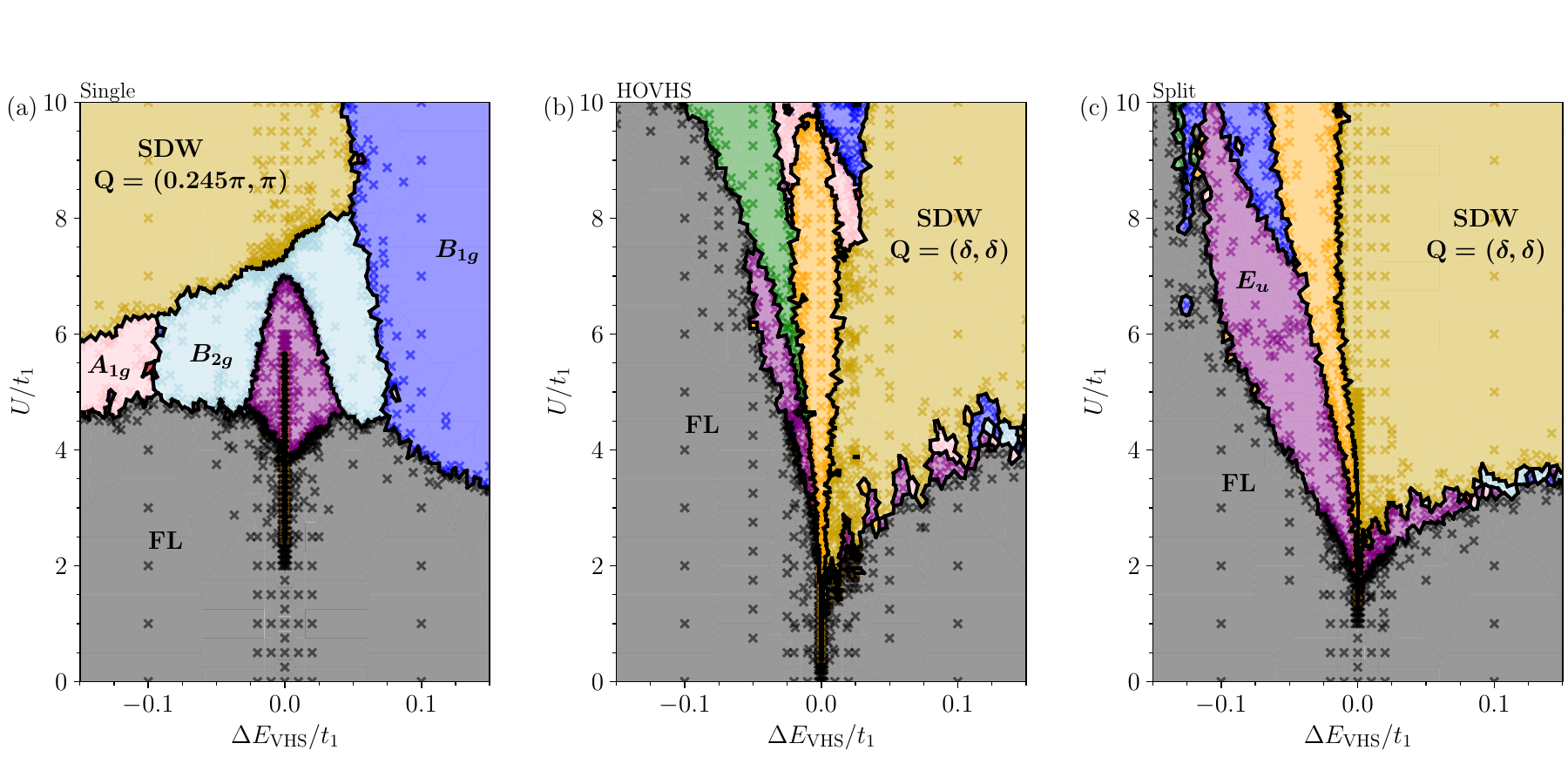}
\captionof{figure}{\centering Doping dependence of the ordered state over a larger range of doping and $U$, in correspondence with Fig.~\ref{fig:doping}. Light blue here is a $B_{2g}$ superconducting state (which is $d_{xy}$-wave, compared to the $B_{1g}$ $d_{x^2-y^2}$-wave symmetry). Note that, in the SDW regions in (b) and (c), $\delta$ attains larger values than in Fig.~\ref{fig:doping}, reaching $\approx \pi/5$.}
\label{fig:doping_largeU}
\end{center}
\twocolumngrid

\end{appendices}

\end{document}